\documentclass[lettersize,journal]{IEEEtran}

\usepackage{amsmath,amsfonts}
\usepackage{algorithmic}
\usepackage{algorithm}
\usepackage{array}
\usepackage[caption=false,font=normalsize,labelfont=sf,textfont=sf]{subfig}
\usepackage{textcomp}
\usepackage{url}
\usepackage{verbatim}
\usepackage{graphicx}
\usepackage{cite}
\usepackage{upgreek}
\usepackage{xcolor}
\usepackage{epstopdf}
\usepackage{dblfloatfix}
\usepackage{hyperref}
\usepackage{hhline}
\usepackage{arydshln}
\usepackage{booktabs}
\usepackage{orcidlink}
\hypersetup{
    colorlinks=true,
    allcolors=blue
}

\begin{document}

\title{Multistatic J-Band Radar TX/RX Chipset in SiGe BiCMOS with Integrated $\times$16 Frequency Multiplier Chain and High EIRP}

\author{Stephan Hauptmeier\,\orcidlink{0009-0008-7924-3457},~\IEEEmembership{Graduate Student Member,~IEEE}, Kennet Braasch\,\orcidlink{0000-0003-0585-3734},~\IEEEmembership{Member,~IEEE}, Till~Ziegler-Bellenberg\,\orcidlink{0009-0006-3720-9889},~\IEEEmembership{Graduate Student Member,~IEEE}, Diana~P.~Cortés~N.\,\orcidlink{0009-0006-7132-1217}, Tobias~T.~Braun\,\orcidlink{0000-0002-9938-5697},~\IEEEmembership{Member,~IEEE}, Michael Höft\,\orcidlink{0000-0001-9352-2868},~\IEEEmembership{Senior Member,~IEEE}, Nils Pohl\,\orcidlink{0000-0001-5362-638X},~\IEEEmembership{Fellow,~IEEE}
\thanks{The research work presented in this paper was funded by the German Research Foundation ("Deutsche Forschungsgemeinschaft") (DFG) under Project-ID 495198532.}
\thanks{Stephan Hauptmeier, Tobias T. Braun and Diana Cortes are with the Institute of Integrated Systems, Ruhr University Bochum, 44801 Bochum, Germany (e-mail: stephan.hauptmeier@rub.de).}
\thanks{Till Ziegler-Bellenberg is with the Fraunhofer Institute,
	FHR, Wachtberg, 53343 North-Rhine Westphalia, Germany.}
\thanks{Kennet Braasch and Michael Höft are with the Department of Electrical and
Information Engineering, Kiel University, 24143 Kiel, Germany.}
\thanks{Nils Pohl is with the Institute of Integrated Systems, Ruhr University
	Bochum, 44801 Bochum, Germany, and also with the Fraunhofer Institute,
	FHR, Wachtberg, 53343 North-Rhine Westphalia, Germany.}    
	\thanks{Spacer for Digital Object Identifier}    
}

\markboth{IEEE Journals,~Vol.~14, No.~8, August~2021}
{Shell \MakeLowercase{\textit{et al.}}: A Sample Article Using IEEEtran.cls for IEEE Journals}


\maketitle

\begin{abstract}
This work presents the design and measurement of a multistatic J-band radar chipset comprising a transmitter and a receiver MMIC, both featuring an integrated $\times$16 frequency multiplier chain for low-frequency local-oscillator distribution and scalable radar configurations.
Multistatic radar architectures can sustain high transmission power and high receiver sensitivity simultaneously, an advantage that is fully leveraged in the present chipset. To this end, a four-way power-combining amplifier chain integrated on the transmitter MMIC delivers an output power of 11.2\,dBm. The resulting measured EIRP is 41\,dBm at 292\,GHz with a collimating PTFE lens and 8.8\,dBm without a lens.
Despite the high frequency-multiplication factor, an on-chip harmonic rejection better than 24\,dBc was measured, while a radiated in-band harmonic rejection of approximately 50\,dBc was achieved through multiple filter stages.
The receiver MMIC incorporates a three-stage low-noise amplifier and exhibits an overall conversion gain of 43.3\,dB at 292\,GHz.
Integrated on-chip patch antennas facilitate system integration and the use of highly directive dielectric lenses, making the chipset suitable for long-range radar measurements, which are demonstrated up to 150\,m.
The MMICs are realized in a 130\,nm SiGe BiCMOS technology, with an f$_\textup{t}$ and f$_\textup{max}$ of 500\,GHz and 610\,GHz, respectively.
\end{abstract}

\begin{IEEEkeywords}
Band-pass filters, BiCMOS integrated circuits, Dielectric lens, Frequency conversion, Frequency multipliers, Harmonic rejection, J-band, Low-noise amplifiers, Millimeter wave radar, Mixers, MMICs, Multistatic radar, Patch antennas, Power amplifiers, Receiver (Rx), Silicon germanium (SiGe), Transmitter (Tx).
\end{IEEEkeywords}

\section{Introduction}
\IEEEPARstart{S}{ub-THz} frequencies have gained significant interest over the last years, as they open up new radar applications in fields of medical screening, imaging, aerosol particle sensing, and non-destructive testing.~\cite{10090458,9855520,8792450,8754785,8516335,8167336,6005348,10934734,8247432,Braasch2026+} 

Even though the large available bandwidth at these frequencies enables high range resolution, a practical deployment of THz radar systems is still hindered by low signal-to-noise ratios (SNR) due to the high path loss, limited transmission power and high receiver noise figure. 
Sub-THz radar systems are therefore typically constrained to short-range applications, highly reflective targets, and limited penetration depth.~\cite{10663766, 8754785} 
To overcome these limitations, recent demonstrators of THz radar applications used high-end measurement devices, such as vector network analyzers (VNA) with frequency extenders, which provide the required dynamic range but are unsuitable for practical applications due to their size and cost.~\cite{10926042,11204970} 
Advances of THz radar systems into practical applications therefore require highly integrated transmitter and receiver monolithic microwave integrated circuits (MMIC) with high output power and low noise figure, respectively. 
\IEEEpubidadjcol
 	\begin{figure}[t]
		\centering
		\includegraphics[width=88.0mm]{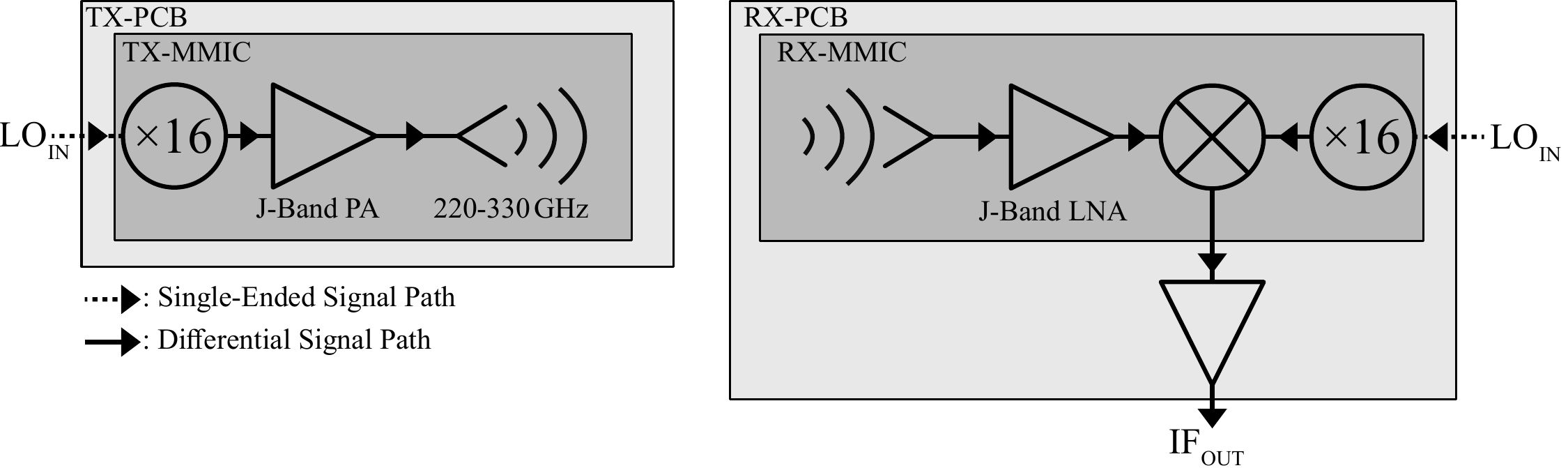}
		\caption{Block diagram of the proposed multistatic radar system, 
		in which both RX and TX feature an integrated $\times 16$ multiplier chain and on-chip antennas, while the TX MMIC incorporates a J-band power amplifier, the RX MMIC includes a J-band LNA and mixer.}
				\label{fig:systemBSB}
	\end{figure}
Silicon germanium (SiGe) BiCMOS technologies have established themselves as a good trade-off between performance and cost, as they provide high f$_\textup{t}$ and f$_\textup{max}$, enabling Sub-THz low noise amplifier (LNA) and power amplifier (PA) design with higher performance compared to CMOS designs, while still offering high integration capabilities and high yield, which is increasing the cost efficiency compared to III-V technologies.~\cite{10979135,B12_Paper,8653991,9499518,10918779}

Within the sub-THz spectrum, the J-band (220\,GHz to 330\,GHz) has attracted particular interest, balancing PA output power, LNA noise performance, and available bandwidth.~\cite{10979135}

A high exploitation of the available J-band bandwidth was demonstrated in \cite{Ziegler_Bellenberg_2025}, where a tuning range of 90\,GHz was achieved, yielding a range resolution of 1.97\,mm. Despite the use of a highly directive lens, the EIRP of only 9.2\,dBm limits the maximum range of the radar system.

In \cite{10147390}, a scalable SiGe radar system is presented, achieving the highest reported EIRP at J-band (27\,dBm at 242\,GHz) with a 3\,dB bandwidth of 50\,GHz. The required silicon lens, however, restricts beamwidth adaptability and increases costs.

		\begin{figure*}[t]
		\centering
		\includegraphics[width=150.0mm]{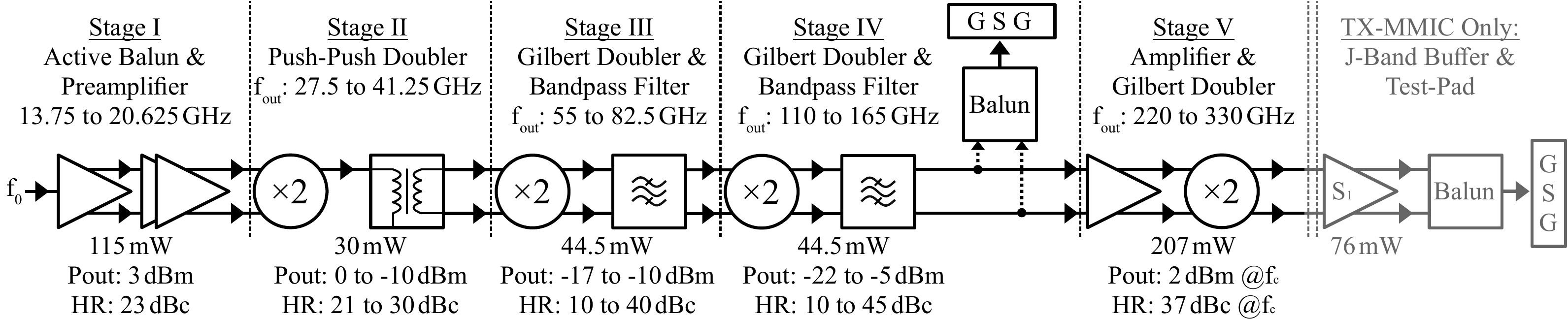}
		\caption{Block diagram of the J-band $\times 16$ multiplier chain with doubler stages, bandpass filter, and amplifier, with simulated output power, DC power consumption, and harmonic rejection (HR) annotated per stage. Measurements can be performed at test pads located at the output of stage IV and at the output of a J-band buffer, the latter being included in the TX-MMIC only.}

		\label{fig:x16_BSB}
		\vspace{-\baselineskip}
	\end{figure*}

This work advances the state of the art by proposing a sub-THz J-band radar system with high EIRP, high receiver sensitivity, and large bandwidth. A multistatic architecture minimizes TX-RX coupling, enabling the use of a high-power J-band PA without receiver saturation. The block diagram of the proposed system is shown in Fig.~\ref{fig:systemBSB}.

To enable spatially separated transmitter and receiver operation with correlated measurements, the LO signal is distributed at low frequencies, minimizing losses and costs in the distribution network.
Both MMICs therefore incorporate an integrated $\times$16 multiplier chain, enabling LO distribution below 21\,GHz. The chain was designed for high harmonic rejection to suppress ghost targets from harmonics near the transmission frequency, despite the multiplier factor of 16.
High intrinsic PA output power and the wide-angle characteristics of the on-chip patch antennas enable large-field-of-view operation, or high-directivity long-range measurements when combined with dielectric lenses.

The proposed MMICs are designed in IHP's 130\,nm SG13G3 SiGe BiCMOS technology, which features heterojunction bipolar transistors (HBT) with $f_\textup{t}$ and $f_\textup{max}$ of 500\,GHz and 610\,GHz, respectively~\cite{IHPG3}.

The following section details the frequency multiplier design. Sections~\ref{Kap:TX} and~\ref{Kap:RX} cover the transmitter and receiver MMIC design with on-wafer and free-space measurements. Section~\ref{Kapitel:System} presents FMCW radar system performance, and Section~\ref{Kapitel:Conclusion} concludes this paper.

\section{Frequency Multiplier Chain Design}
\label{Kap:X16}

The proposed frequency multiplier chain consists of a cascade of four doubler stages, with additional bandpass filters and amplifiers between the stages to increase the harmonic rejection and to compensate for the losses along the chain. The block diagram of the multiplier chain is shown in Fig.~\ref{fig:x16_BSB}. The entire chain is further separated into 5 stages, for the active balun and the preamplifier, and the four doubler stages.

\begin{figure}[t]
		\centering
		\includegraphics[width=80.0mm]{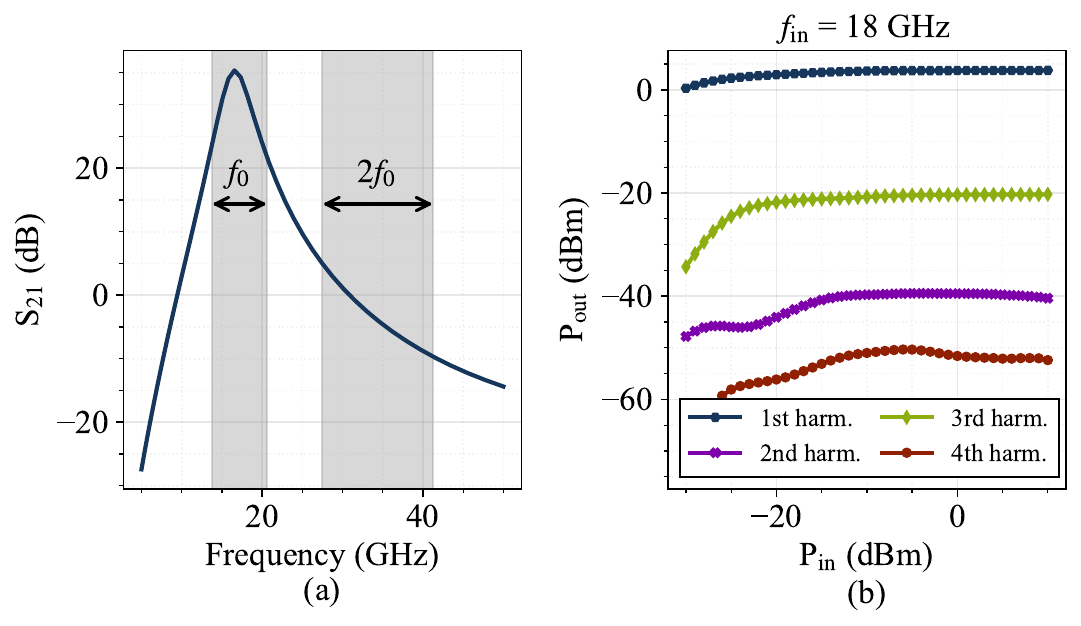}
		\caption{Simulated $S_{21}$ (a) and output power over input power (b) of Stage I.}

		\label{fig:Simc_S1}
	\end{figure}

The first stage consists of an active balun and two preamplifier stages. The active balun generates the differential signal with lower area consumption than passive alternatives. The entire stage achieves a gain of up to 35\,dB at the fundamental frequency, as shown by the simulated $S_{21}$ and output power in Fig.~\ref{fig:Simc_S1}.
The high gain and early saturation of the first stage minimize the impact of input power variations, making the system robust against LO distribution losses that arise when driving multiple transceivers from a single source. The LC loads of all Stage I components create a sharp gain peak, suppressing undesired harmonic generation. Furthermore, a $S_{21}$ below 0\,dB across most of the $2f_0$ bandwidth ensures rejection of second harmonic content in the LO input.

	\begin{figure}[t]
	\centering
	\includegraphics[width=55.0mm]{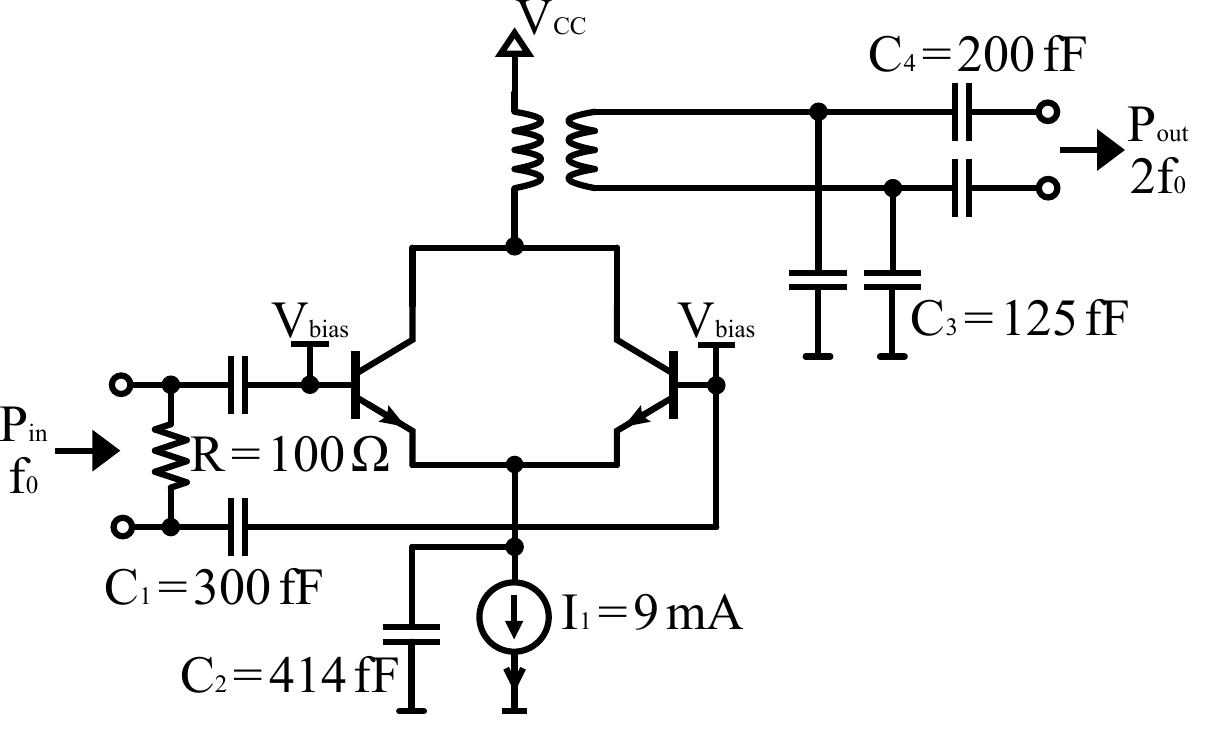}
	\caption{Schematic of the Stage II push-push doubler with transformer load.}

	\label{fig:Simc_S2_Schematic}
	\vspace{-\baselineskip}
	\end{figure}

	The second stage employs a push-push doubler with a transformer load, shown in Fig.~\ref{fig:Simc_S2_Schematic}. At these low input frequencies, the push-push topology is more compact than a bootstrapped Gilbert cell, which requires $\lambda/4$ transmission lines. The transformer load simultaneously acts as a balun, compensating for the single-ended output.

	The push-push doubler input is matched with a parallel 100\,$\Omega$ resistor, which ensures that the sharp gain peak of the first stage is not affected by the doubler input impedance, without using space-intensive reactive matching networks.

		\begin{figure}[t]
		\centering
		\includegraphics[width=70.0mm]{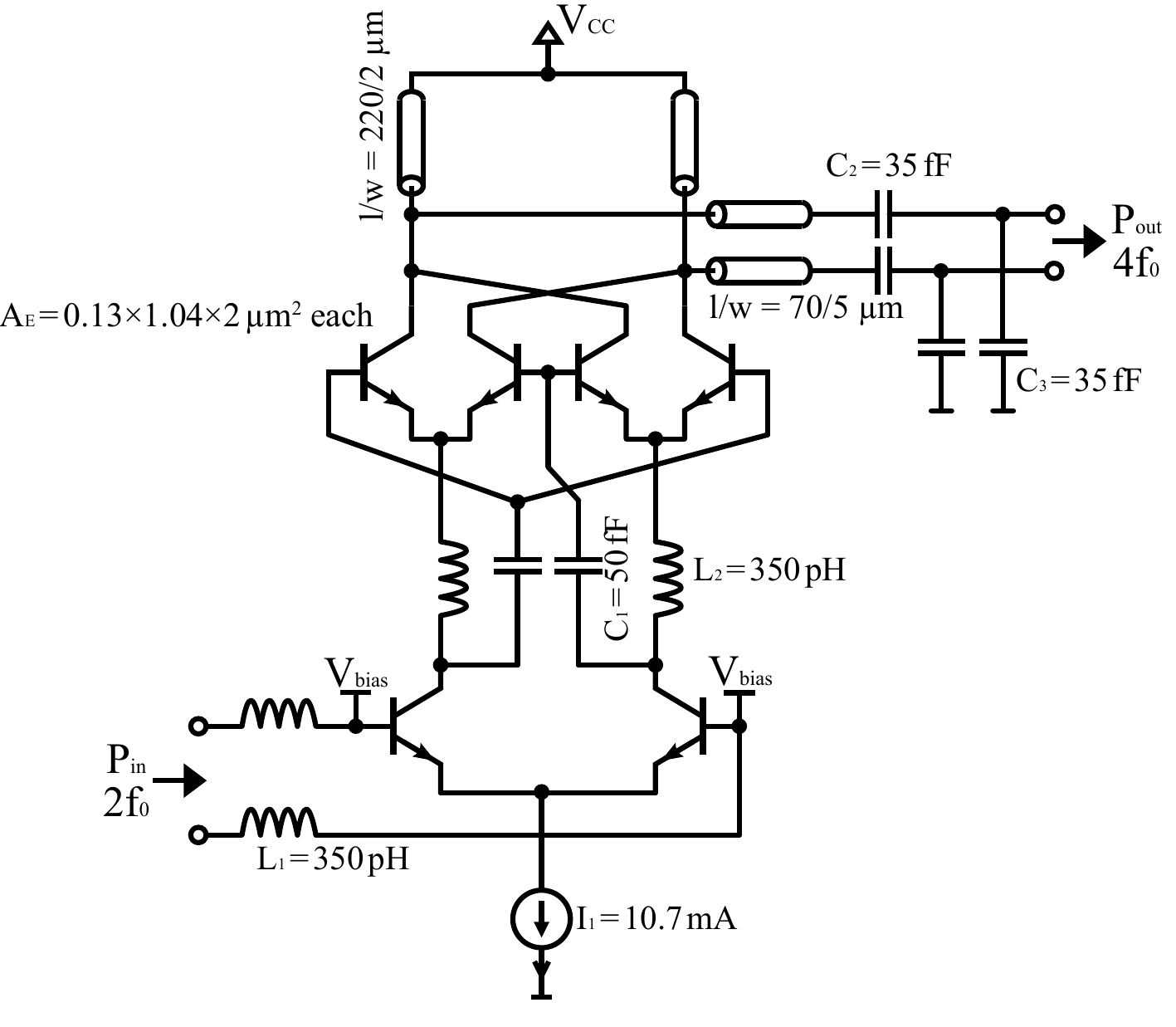}
		\caption{Schematic of the Stage III Gilbert cell-based frequency doubler.}

		\label{fig:Simc_S3_Schematic}
	\end{figure}
					\begin{figure}[t]
		\centering
		\includegraphics[width=85.0mm]{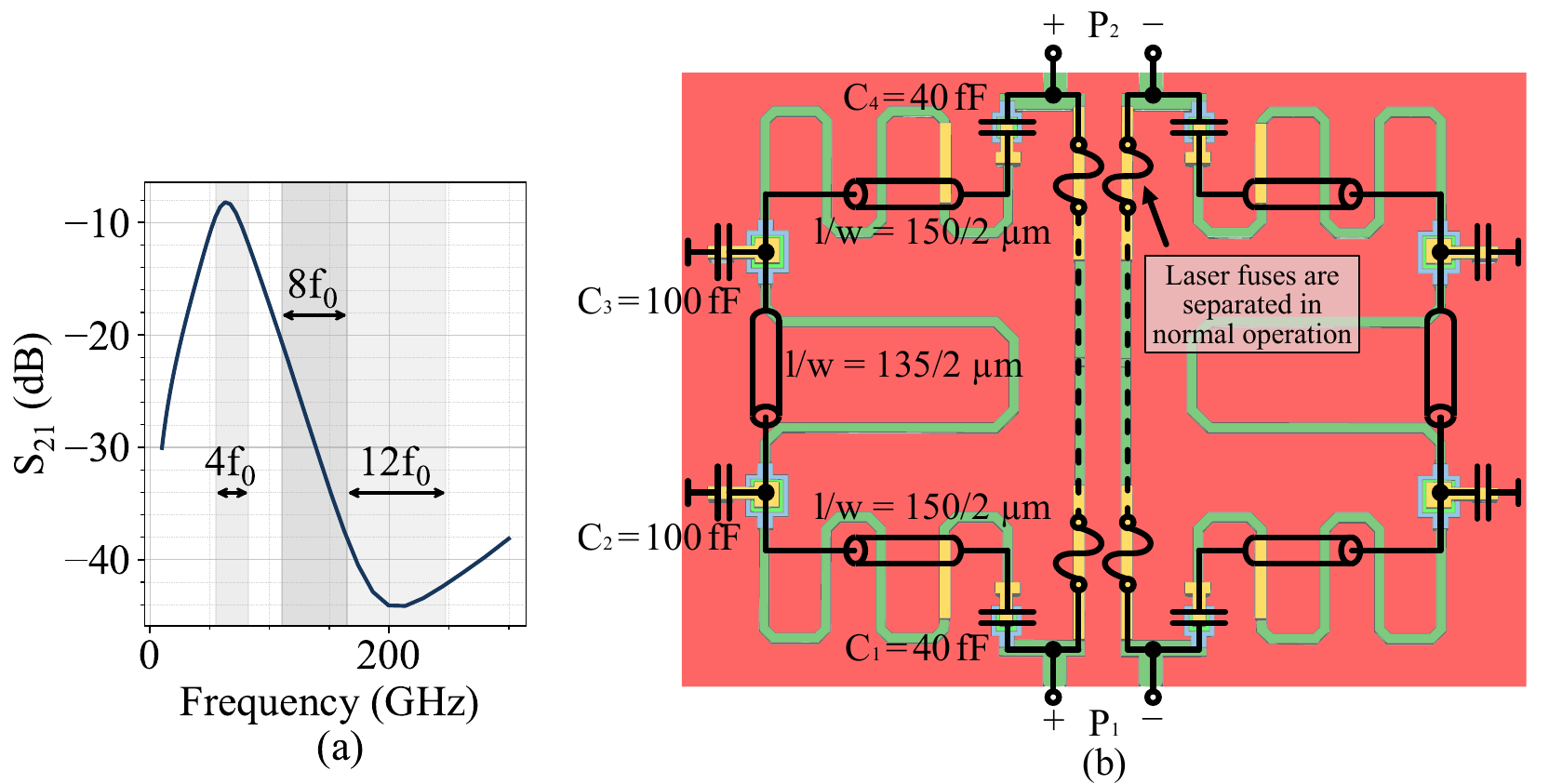}
		\caption{Simulated $S_{21}$ (a) and layout with overlaid schematic (b) of the Stage III filter.}

		\label{fig:S3_Filter}
	\end{figure}

		At the elevated operating frequencies of Stage III, a Gilbert cell topology is adopted for the doubler. Although $\lambda/4$ transmission lines would still reach $1.3\,mm$ in length, inductors provide a compact alternative that substantially improves doubler gain and reduces amplitude imbalance, despite not delivering the full 90$^\circ$ phase shift required for ideal bootstrapped operation.

		The third-stage doubler output is followed by a bandpass filter, whose transmission characteristics and layout are shown in Fig.~\ref{fig:S3_Filter}. The filter is implemented with series transmission lines and capacitors, as well as capacitors to ground. Despite a minimum insertion loss of 8\,dB, the filter substantially improves harmonic rejection, attenuating the 8th and 12th harmonics by 20\,dB to 44\,dB.

	Stage IV comprises the third doubler, realized as a bootstrapped Gilbert cell, and a second bandpass filter. The circuit topology follows Fig.~\ref{fig:Simc_S3_Schematic}, with $\frac{\lambda}{4}$ transmission lines replacing inductors and element dimensions scaled to the higher operating frequencies. The second filter adds shorted transmission lines to increase suppression of low-order harmonics that could otherwise be upconverted to the target band at nonlinear devices.

 	\begin{figure}[t]
 		\centering
 		\includegraphics[width=85.0mm]{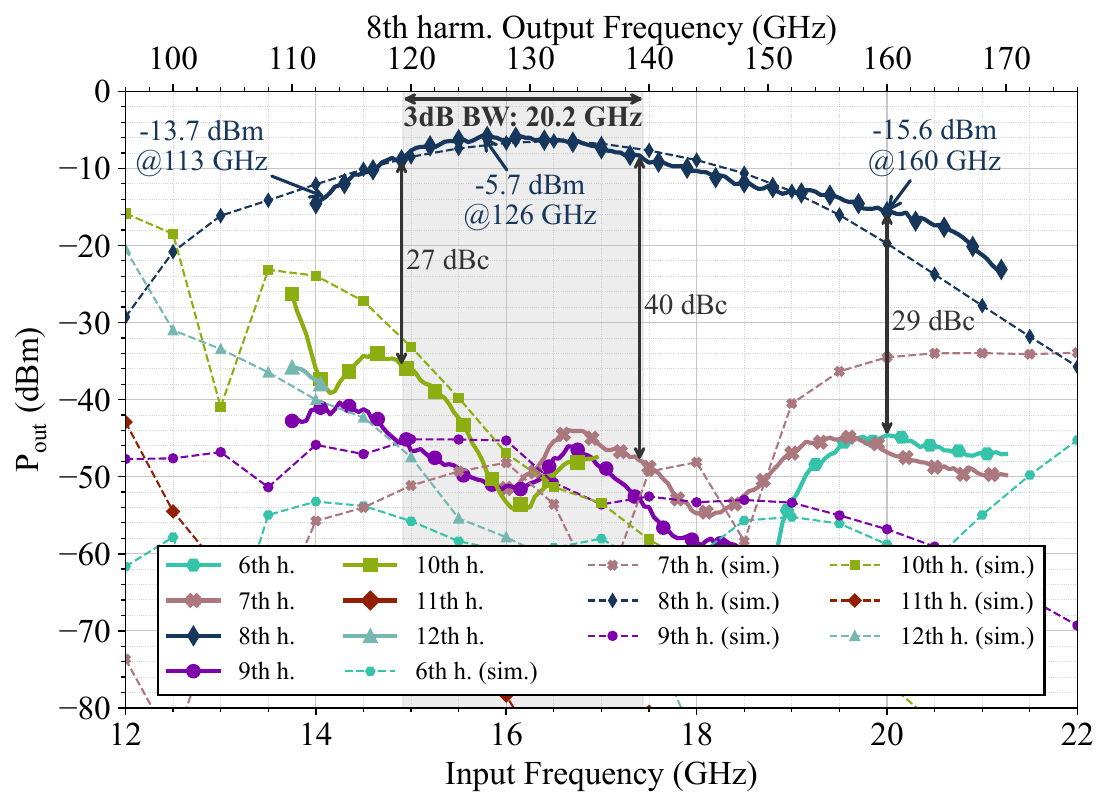}
 		\caption{Measured and simulated output spectrum of stage IV at the test pad. Measurements and simulation include the losses of the balun and the pad.}

 		\label{fig:S4_Sim_Chain}
 	\end{figure}
\begin{figure}[t]
 		\centering
 		\includegraphics[width=80.0mm]{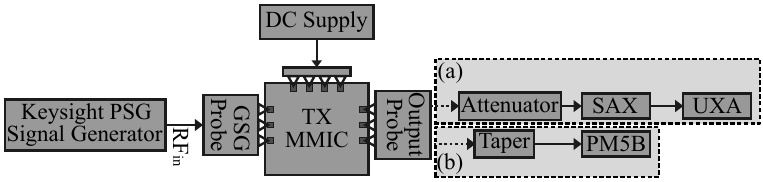}
		\caption{Measurement setups for on-chip spectrum (a) and power measurements (b). In all presented results, losses of probes, mixer, attenuators, taper, and cables are de-embedded.}

 		\label{fig:Meas_Setup}
 		\vspace{-\baselineskip}
 	\end{figure}

	A fusable connection to a test pad is integrated on the transmitter MMIC to measure the output signal of stage IV. The measurement setup consists of a VDI WR6.5 spectrum analyzer extender (SAX) with a Keysight UXA, as shown in Fig.~\ref{fig:Meas_Setup}. The input signal was generated by a Keysight PSG signal generator with an output power of 0\,dBm. The measured harmonics are shown in Fig.~\ref{fig:S4_Sim_Chain} together with the simulated output harmonics. Losses of the measurement equipment are de-embedded, however, the presented results contain pad and balun losses. 
	The measured 8th harmonic shows a good agreement with the simulated results, with a maximum measured power of -5.7\,dBm, a 3\,dB bandwidth of 20.2\,GHz and a harmonic rejection better than 27\,dBc to inband harmonics within the 3\,dB bandwidth. As expected from the filter characteristics, the available bandwidth is constrained at this point of the multiplier chain, however this will be compensated by a D-band buffer at the input of stage V.

	 	 	 		\begin{figure*}[t]
 		\centering
 		\includegraphics[width=180.0mm]{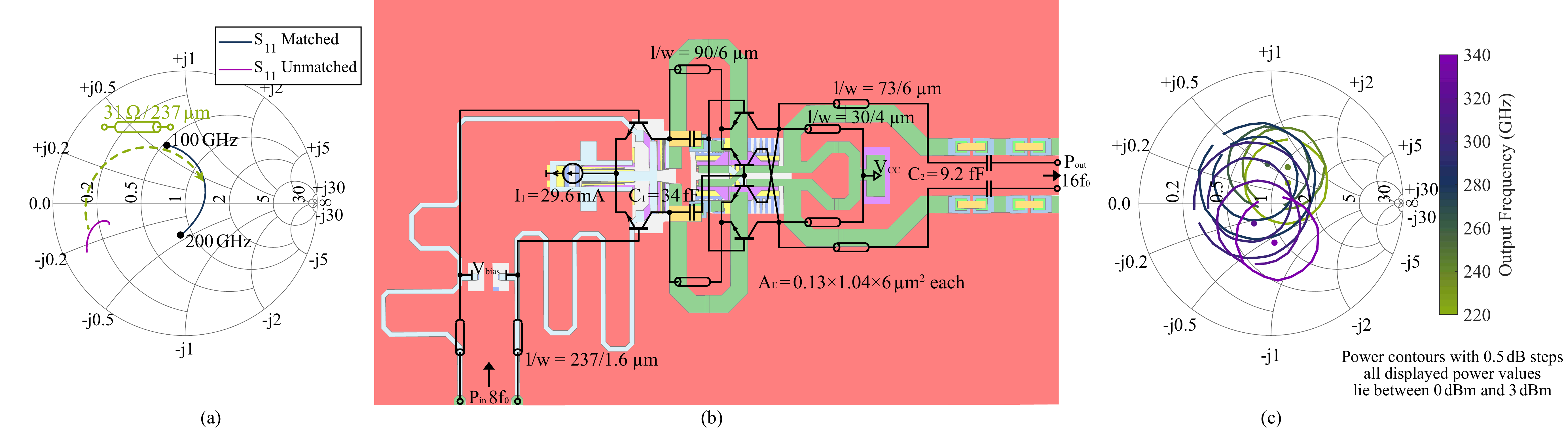}
 		\caption{Stage V input matching shown in the Smith chart (a). Layout and schematic of the stage V doubler (b).
		Simulated power contours of the doubler, showing the output matching to a reference impedance of 100\,$\Omega$ (c).}

 		\label{fig:S5_SchematicX2}
 		\vspace{-\baselineskip}
 	\end{figure*}

 	The bootstrapped Gilbert cell doubler used as the final J-band doubler stage was partially presented in~\cite{11103985}. Fig.~\ref{fig:S5_SchematicX2} provides a more detailed design overview, covering the circuit layout and matching networks. A single series transmission line achieves broadband input matching, visible in the $S_{11}$ Smith chart, and the output power contours confirm good matching around 100\,$\Omega$ with only 0.5\,dB variation from 220\,GHz to 340\,GHz. As shown in~\cite{11103985}, this topology achieves moderate conversion loss and reduces the requirements on the D-band driver, lowering DC power consumption while leaving output power to be increased by the subsequent J-band amplifier stage.

	To verify the functionality of the entire multiplier chain, the output signal was measured at a test pad on the TX-MMIC\@. The test pads lie after a J-band buffer amplifier stage, which is not part of the multiplier chain, as it was not used at the RX-MMIC\@. However, these pads allow the best possible verification of the multiplier chain performance, regarding bandwidth, output power and harmonic rejection. Measurements were performed with the setup presented in Fig.~\ref{fig:Meas_Setup}.

	As expected from the broadband doubler and D-band buffer design, the bandwidth limitation of Stage IV is compensated in Stage V, yielding a measured 3\,dB bandwidth of 77.7\,GHz around 288\,GHz. The maximum output power of the 16th harmonic is $-$1.5\,dBm, with spectrum and power measurements in excellent agreement. Harmonic rejection exceeds 11\,dBc across the full WR3.4 waveguide bandwidth and 26\,dBc within the 3\,dB bandwidth.
	\begin{figure}[t]
 		\centering
 		\includegraphics[width=88.0mm]{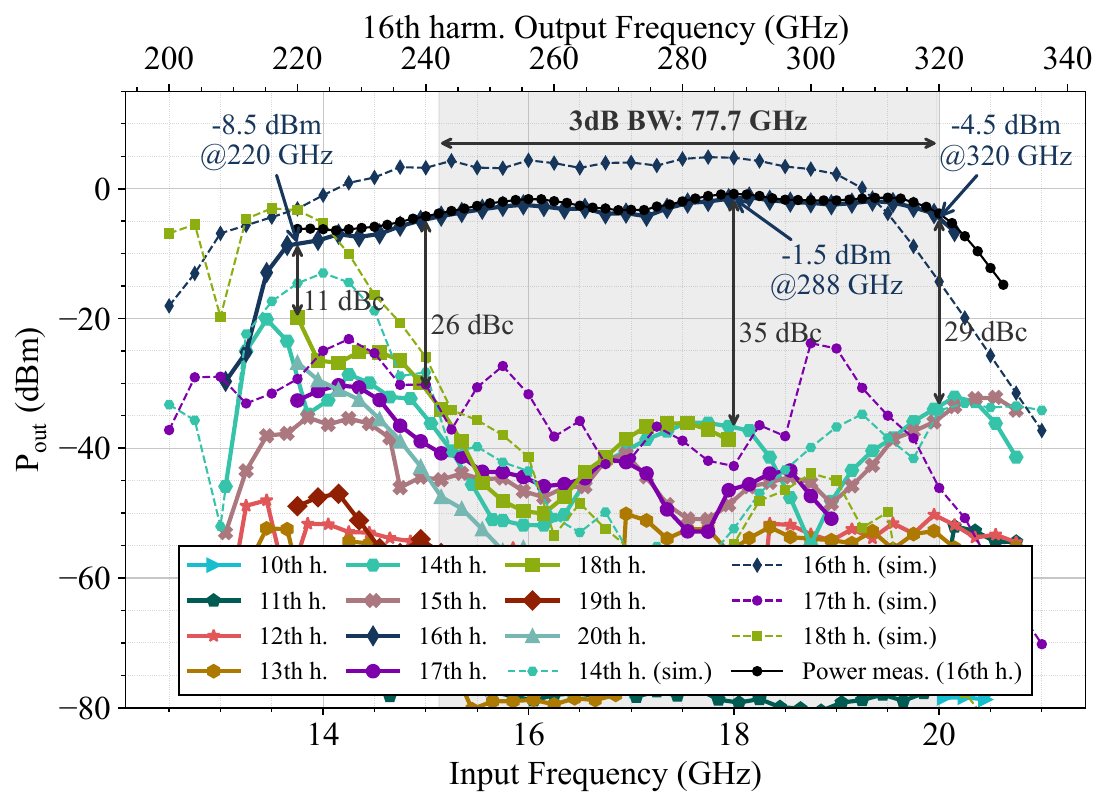}
 		\caption{Measured and simulated output harmonics and power at the output of Stage V with an additional J-band buffer stage. The input power equals 0\,dBm. All results include the losses of the balun and the pad.}

 		\label{fig:S5_Sim_Chain}
 		\vspace{-\baselineskip}
 	\end{figure}

	Compared with simulations, a slight upward frequency shift is observed, which could first be observed at the Stage IV output. The deviation in 16th harmonic output power originates from layout parasitics in the J-band doubler and amplifier, with discrepancies already visible at the doubler breakout and accumulating across subsequent stages. The balun may further introduce unexpected losses through load-pulling effects caused by input impedance variations at the amplifier output.
	The simulation accuracy for undesired harmonics is limited, as they may arise from higher-order intermodulation products sensitive to the  limited bandwidth of S-parameter files. Given the system complexity and multiplication factor of 16, the overall simulation-measurement agreement is still considered good.

	Compared to the state of the art in Table~\ref{TabelleMultiplier}, the proposed multiplier chain offers excellent broadband performance and high harmonic rejection. The higher DC power consumption results from multiple amplifier stages required to maintain high gain and compensate for bandpass filter losses.

\begin{table}[ht!]
	\centering
	\caption{Comparison of on-chip measured state-of-the-art J-band frequency multipliers}
	\label{TabelleMultiplier}

		\centering
\setlength{\tabcolsep}{5pt}
\begin{tabular}{cccccccc}
\textbf{Ref.}   & \textbf{\begin{tabular}[c]{@{}c@{}}f$_{\textup{t}}$/f$_{\textup{max}}$\\ (GHz)\end{tabular}} & \textbf{Type}       & \textbf{\begin{tabular}[c]{@{}c@{}}f$_\textup{c}$\\ (GHz)\end{tabular}} & \textbf{\begin{tabular}[c]{@{}c@{}}P$_\textup{Sat}$\\ (dBm)\end{tabular}} & \textbf{\begin{tabular}[c]{@{}c@{}}BW$_{\textup{3dB}}$\\ (GHz)\end{tabular}} & \textbf{\begin{tabular}[c]{@{}c@{}}$\boldsymbol{\mathrm{HR}_{f_\mathrm{c}}}$\\ (dBc)\end{tabular}} & \textbf{\begin{tabular}[c]{@{}c@{}}P$_{\textup{DC}}$\\ (mW)\end{tabular}} \\ \hline
\cite{11234660} & 300\,/\,500   & $\times$8           & 290                                                                     & -0.5                                                                      & 71                                                                                 & $>$18                                                                 & 122                                                              \\
\cite{10817731} & 470\,/\,650   & $\times$6           & 258                                                                     & 4.3                                                                       & 40                                                                                 & $>$30$^a$                                                             & 119                                                              \\
\cite{10867300} & 300\,/\,500   & $\times$8           & 320                                                                     & 2                                                                         & 40                                                                               & 29                                                                  & 509                                                              \\
\cite{10438581} & 470\,/\,650   & $\times$9           & 270                                                                     & 9.6                                                                       & 56                                                                                & $>$40                                                                 & 660                                                              \\
\cite{9732898}  & 300\,/\,500   & $\times$18          & 240                                                                     & 6.2                                                                       & 41                                                                                 & 35$^{a,b}$                                                           & 429                                                              \\ \hline
\textbf{This}   & \textbf{500\,/\,610} & \textbf{$\times$16} & \textbf{288}                                                            & \textbf{-1.5}                                                             & \textbf{78}                                                                         & \textbf{35}                                                          & \textbf{511$^c$}                                                   \\ \hline
\end{tabular}

		\footnotesize{$^a$\,estimated from graph, $^b$simulated, $^c$ measured P$_\textup{DC}$ of entire TX MMIC, with subtracted simulated DC power of the unused PA stages.}

\end{table} 

\begin{figure*}[t]
 		\centering
 		\includegraphics[width=175.0mm]{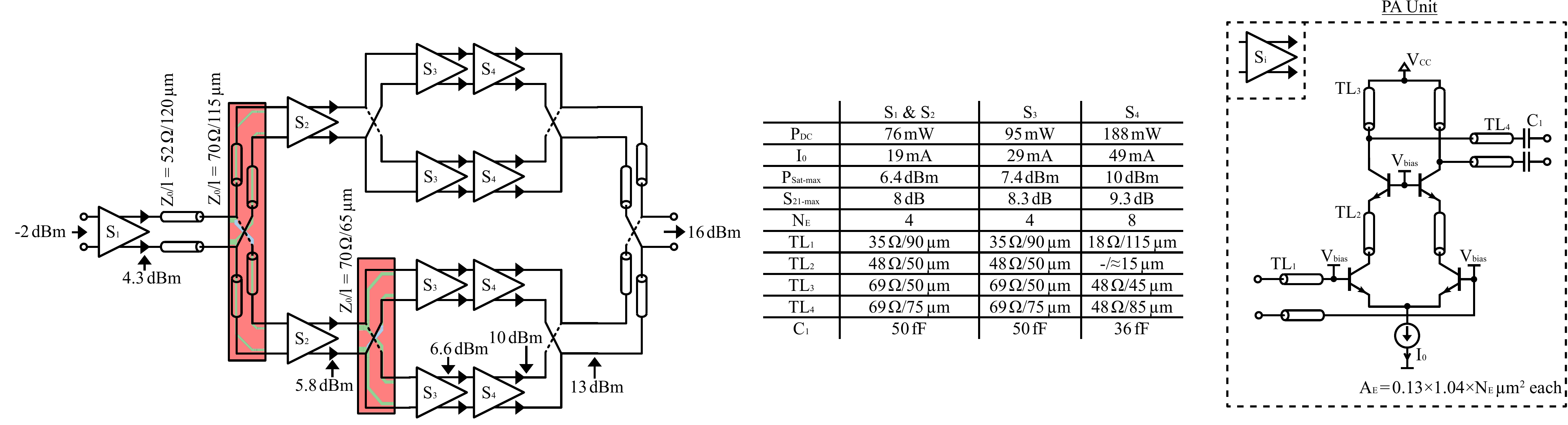}
 		\caption{Block diagram of the four-stage J-band power amplifier with transmission-line-based power splitting and combining network (a), the individual design parameters of each stage (b), and a representative single-stage schematic (c). Annotated power levels are simulated values.}

 		\label{fig:Amp_BSB}
 		\vspace{-\baselineskip}
 	\end{figure*}

\section{Transmitter MMIC}
\label{Kap:TX}
\begin{figure}[t]
	\centering
	\includegraphics[width=70.0mm]{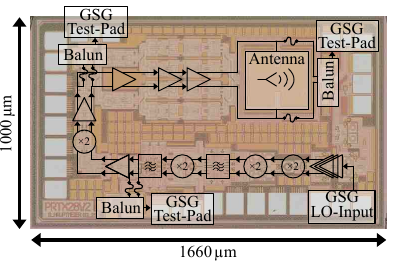}
	\caption{Micrograph of the TX MMIC with overlaid block diagram.}

	\label{fig:TX_pic}
	\vspace{-\baselineskip}
\end{figure}
Since the $\times 16$ multiplier chain output power is insufficient for long-range radar applications, a J-band power amplifier is integrated on the transmitter MMIC.

The PA design shown in Fig.~\ref{fig:Amp_BSB} comprises four stages (S$_1$ to S$_4$) with a transmission-line-based power dividing and combining network. While the first three stages prioritize gain over output power, the final stage maximizes output power by employing 8 transistor unit cells biased slightly above the f$_\textup{T}$-optimal current. The incorporation of tail current sources in all stages ensures robust operation against temperature variations caused by the high power dissipation of the power-combining stages.

The complete transmitter MMIC is shown in Fig.~\ref{fig:TX_pic}. To enable individual verification of the multiplier chain and power amplifier, test pads are integrated at the output of the octupler, at the output of the PA stage S$_1$, and at the final transmitter output. Each test pad is preceded by a balun, enabling reliable harmonic measurements, as certain harmonics appear as common-mode and others as differential-mode signals. Laser fuses allow the test pads to be disconnected during normal operation, or permit isolation of the antenna or the subsequent transmitter chain when the test pads are used for on-wafer measurements.

\begin{figure}[t]
 		\centering
 		\includegraphics[width=88.0mm]{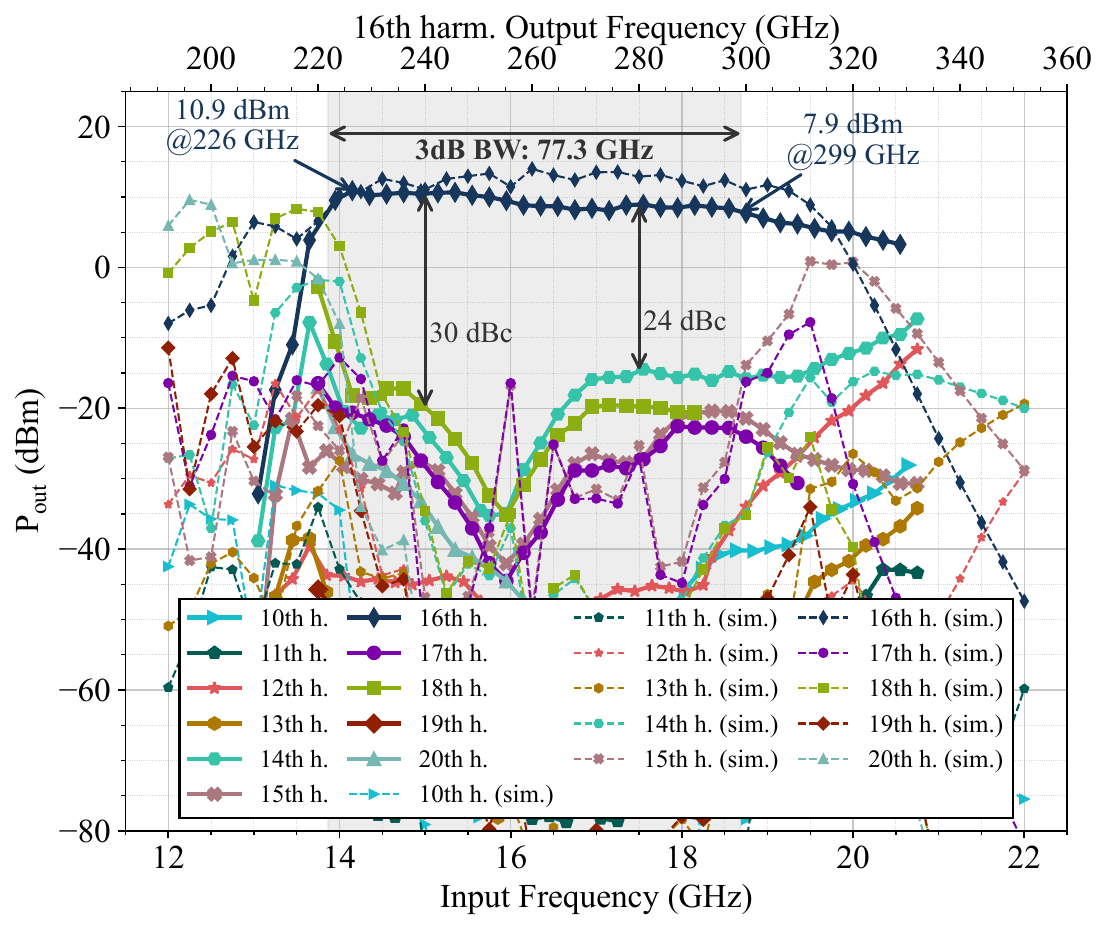}
		\caption{On-chip measured and simulated output harmonics of the entire transmitter chain, measured with a WR3.4 SAX. Over the 77.3\,GHz 3\,dB bandwidth, the 16th harmonic dominates and harmonic rejection exceeds 24\,dBc over most of the bandwidth. For input frequencies below 13.75\,GHz, a WR5.4 SAX was used.}

 		\label{fig:Meas_HarmonicsPlot}
 		\vspace{-\baselineskip}
 	\end{figure}

On-wafer measurements of the TX MMIC were performed using the setup shown in Fig.~\ref{fig:Meas_Setup}. Results are presented in Fig.~\ref{fig:Meas_HarmonicsPlot}, where all measured output harmonics are plotted. The harmonic rejection exceeds 24\,dBc across most of the 3\,dB bandwidth, with the 14th, 15th, 17th, and 18th harmonics being the most significant spurious components, consistent with simulation. Compared to the multiplier chain output, the harmonic rejection is reduced, as these harmonics fall within the bandwidth of the power amplifier and, since the 16th harmonic may drive the amplifier into compression, lower-power harmonics experience higher gain than the 16th harmonic.

\begin{figure}[t]
 		\centering
 		\includegraphics[width=88.0mm]{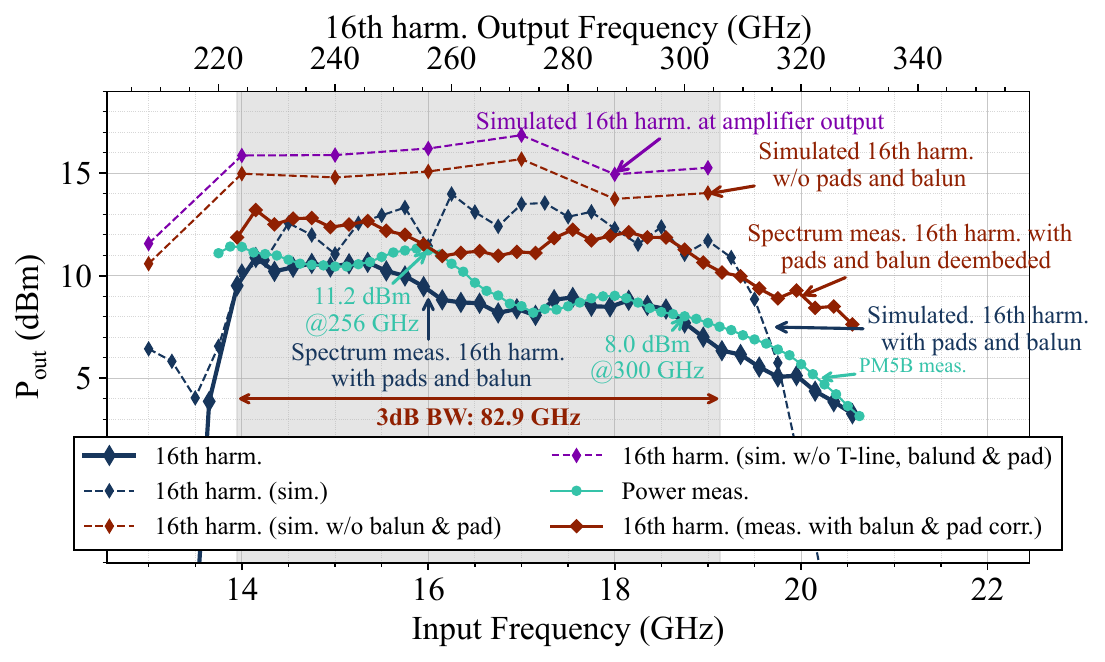}
		\caption{Measured and simulated 16th harmonic output power of the transmitter chain. Results from a spectrum analyzer and a power meter are included. The spectrum analyzer trace is also corrected for balun and pad insertion losses measured in a back-to-back configuration. Simulated results refer to the PA output, balun input, and pad.}

 		\label{fig:Meas_16thHarmPlot}
 		\vspace{-\baselineskip}
 	\end{figure}

Fig.~\ref{fig:Meas_16thHarmPlot} shows a detailed view of the 16th harmonic output power. The 16th harmonic output power measured with the SAX and with the PM5B power meter is plotted. Both curves show good agreement. Differences may be caused by different probes used for the two measurements.
For the SAX measurement curve, also a curve with a correction of the back-to-back measured losses of the balun and the pad is shown.
This allows a more precise estimation of the on-chip available power and bandwidth at the antenna input. 
This curve shows a 3\,dB bandwidth of 82.9\,GHz with a maximum output power of 13.2\,dBm. 
Compared to simulations, variations are less than 5\,dB which can be considered good agreement, regarding possible error propagation of the multiple stages in the TX chain.  

 	 	 	 	\begin{figure}[t]
 		\centering
 		\includegraphics[width=88.0mm]{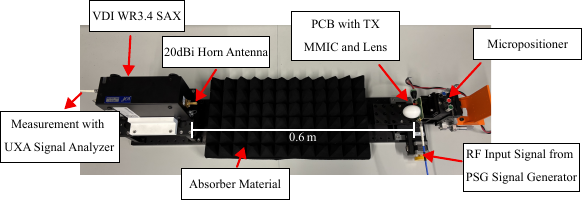}
		\caption{EIRP measurement setup using the SAX and UXA\@. Free-space path loss, mixer loss, and horn antenna gain are de-embedded from all results.}

 		\label{fig:Meas_Setup_EIRP}
 		\vspace{-\baselineskip}
 	\end{figure}

		\begin{figure}[t]
 		\centering
 		\includegraphics[width=88.0mm]{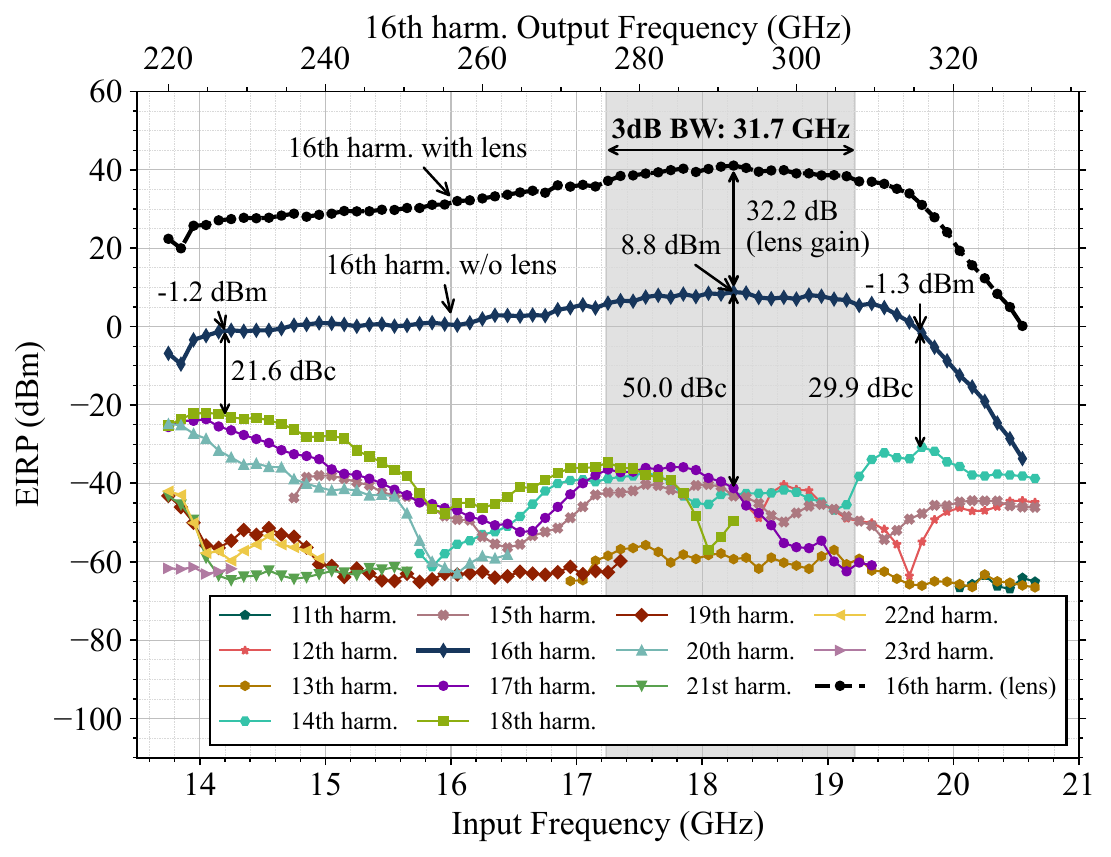}
		\caption{TX system EIRP measured with a WR3.4 SAX. All harmonics are captured without lens, and the 16th harmonic is additionally shown with lens.}

 		\label{fig:Meas_Freespace}
 	\end{figure}
		\begin{figure}[t]
 		\centering
 		\includegraphics[width=88.0mm]{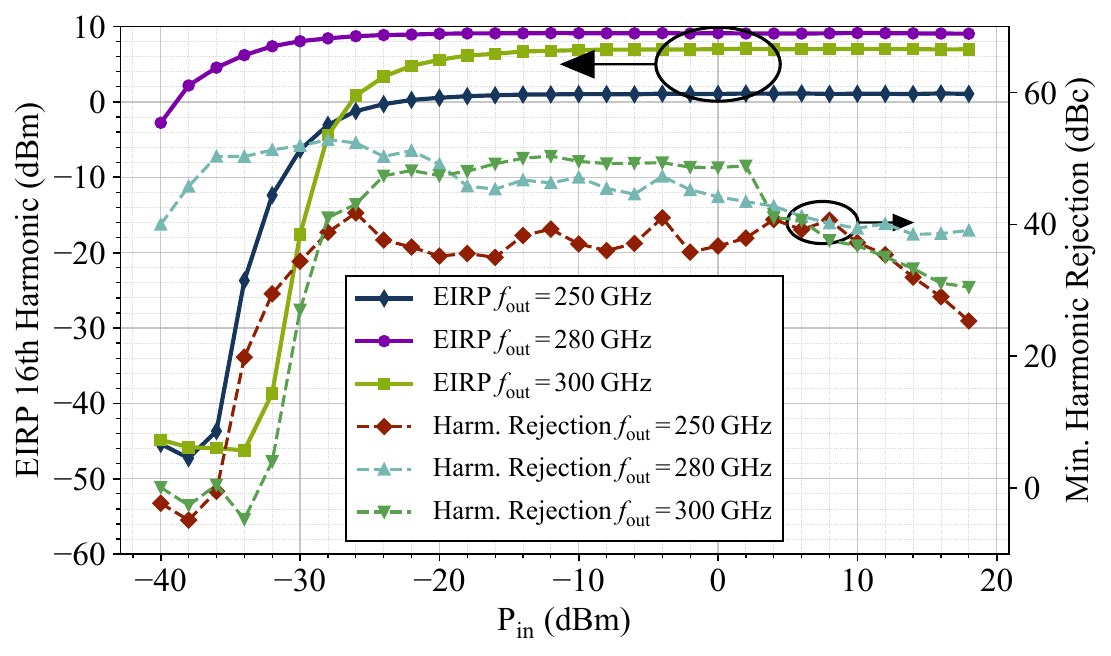}
		\caption{Measured 16th harmonic EIRP and minimum harmonic rejection over the fundamental input power. Measurements were performed with a freespace setup without lens.}

 		\label{fig:Meas_Freespace_PoutVsPin}
 	\end{figure}

A differential rectangular on-chip patch antenna, integrated on both MMICs, is realized in the top metal layer over an on-chip ground plane. The 252\,$\mu$m $\times$ 280\,$\mu$m patch resonates at 280\,GHz, with a maximum efficiency of -6\,dB and a simulated directivity of approximately 6\,dBi.     
An EIRP measurement of the TX was performed with the setup shown in Fig.~\ref{fig:Meas_Setup_EIRP}. Corresponding results are shown in Fig.~\ref{fig:Meas_Freespace}, where the EIRP of all TX harmonics is plotted. A significant improvement of the harmonic rejection can be observed, caused by the limited bandwidth of the utilized patch antenna. Also, perfect destructive superposition due to the symmetric patch antenna increases the harmonic rejection for common-mode harmonics.

By placing the antenna's resonance frequency in the upper half of the J-band, the decreasing PA output power gets partly compensated by the increasing antenna gain, resulting in a 3\,dB bandwidth of 31.7\,GHz and 10\,dB EIRP variation from 227\,GHz to 315\,GHz. The maximum EIRP is measured to be 8.8\,dBm without lens at 292\,GHz, which can be significantly increased to 41\,dBm with a PTFE lens, based on the design in \cite{6459227}, with a diameter of 36\,mm.

Fig.~\ref{fig:Meas_Freespace_PoutVsPin} shows the measured 16th harmonic EIRP and minimum harmonic rejection vs.\ input power. Due to the high gain and early saturation of the first multiplier stage, both quantities remain nearly constant from -25\,dBm to 0\,dBm, confirming robustness against input power variations.

\section{Receiver MMIC}
\label{Kap:RX}
\begin{figure}[t]
	\centering
	\includegraphics[width=70.0mm]{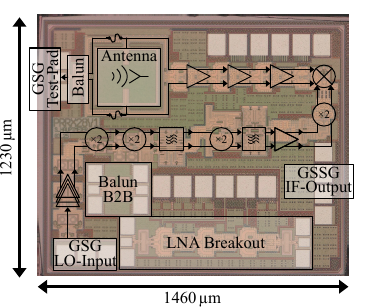}
	\caption{Micrograph of the RX MMIC with overlaid block diagram.}

	\label{fig:RX_pic}
	\vspace{-\baselineskip}
\end{figure}

 	\begin{figure}[t]
 	\centering
 	\includegraphics[width=85.0mm]{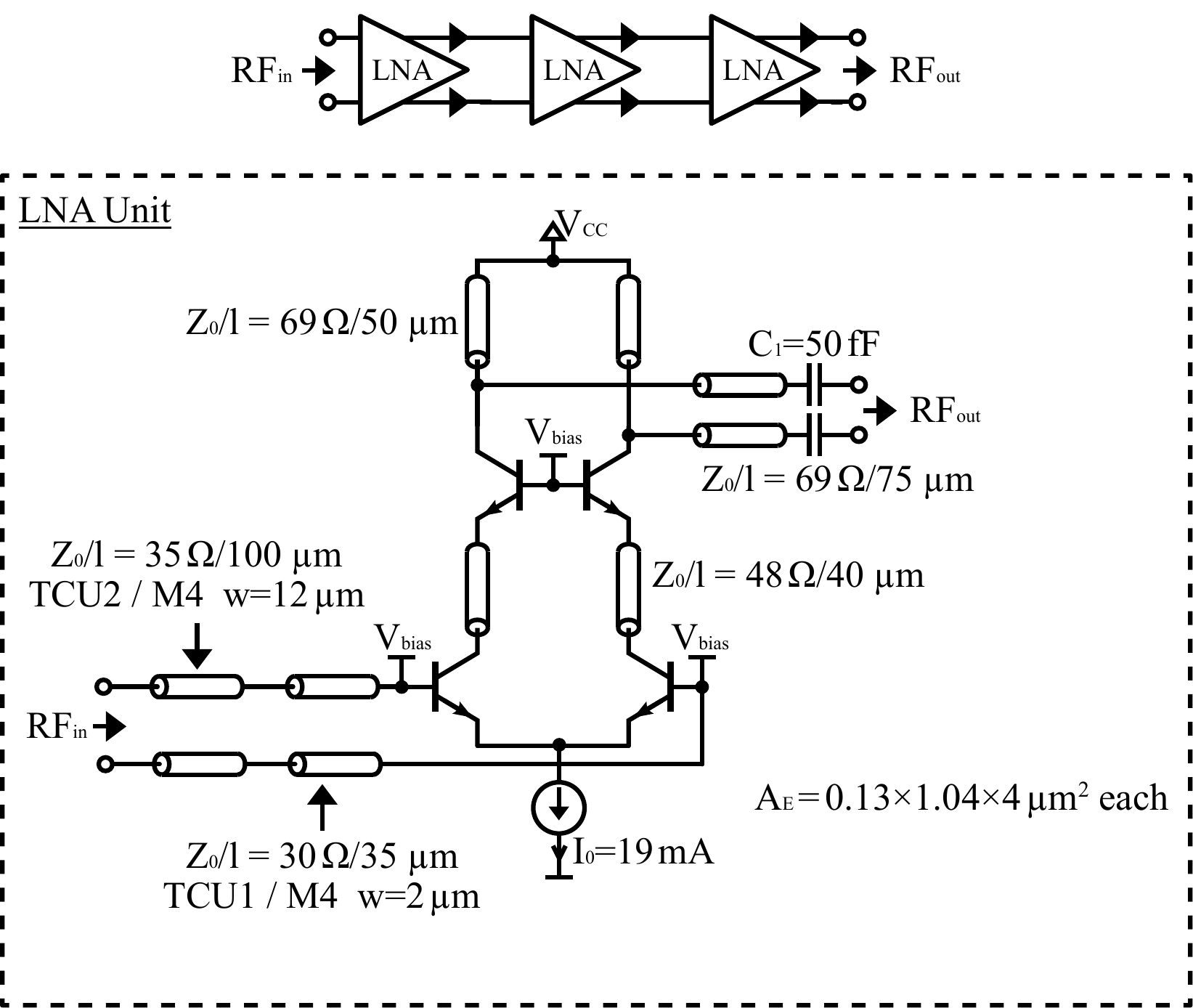}
 	\caption{Block diagram of the 3-stage LNA and circuit diagram of the single LNA unit.}

 	\label{fig:LNA_Schematic}
 	\end{figure}

\begin{figure}[t]
 		\centering
 		\includegraphics[width=65.0mm]{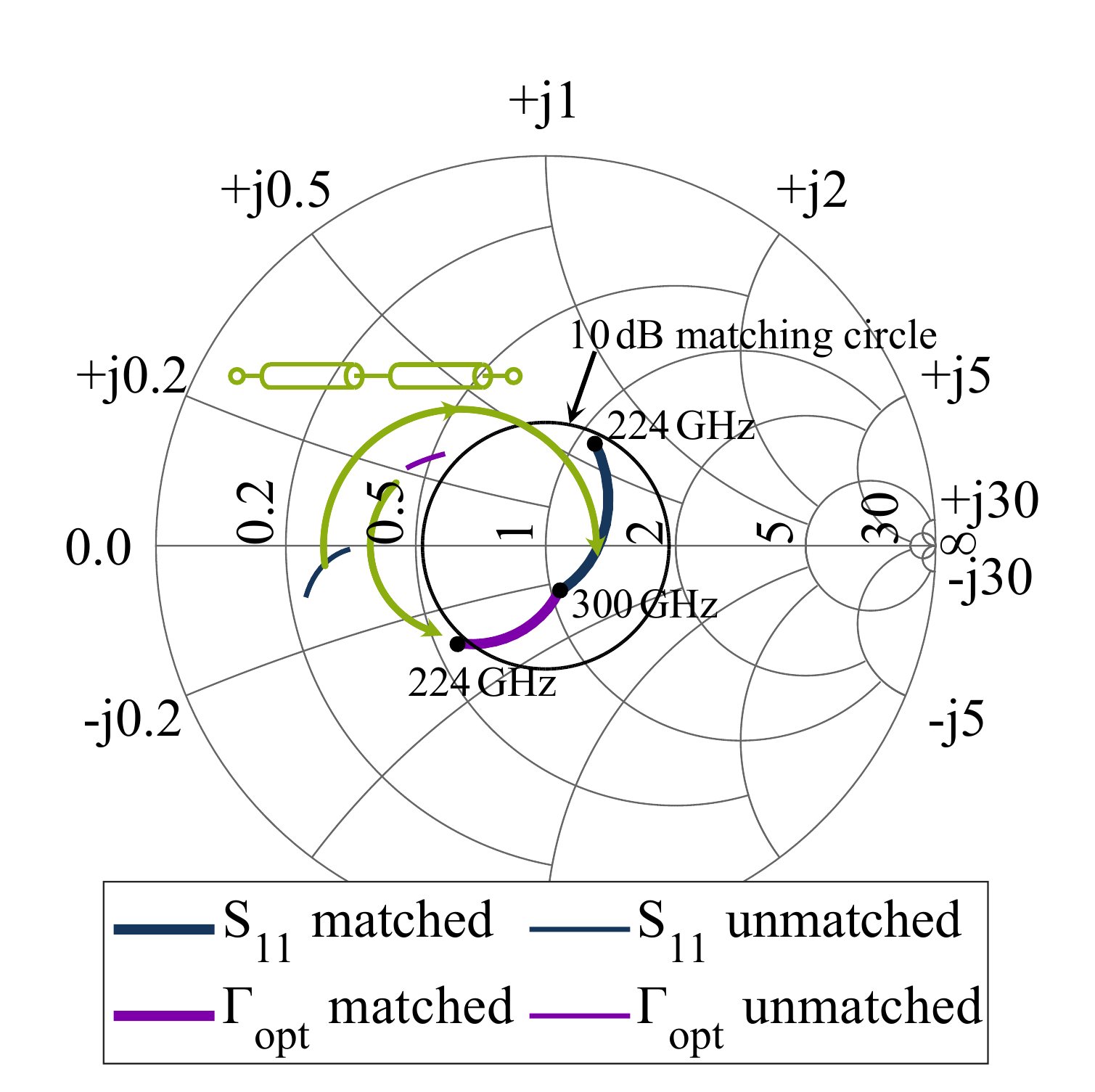}
 		\caption{Smith chart showing the simulated LNA input and noise reflection coefficients with and without the matching network. The series transmission line achieves simultaneous matching of both coefficients across the band.}

 		\label{fig:LNA_Matching}
 	 \end{figure}

The receiver MMIC, which is shown in Fig.~\ref{fig:RX_pic}, includes the $\times 16$ frequency multiplier chain to generate the LO signal for a fundamental down-conversion mixer.
The RF signal is received by an on-chip antenna, similar to the TX antenna, and amplified by a three-stage LNA.
Block and circuit diagrams of the LNA are shown in Fig.~\ref{fig:LNA_Schematic}.
A series transmission line input matching network enables simultaneous optimization of noise figure and input return loss, as shown in Fig.~\ref{fig:LNA_Matching}. These two parameters typically move in opposing directions in the Smith chart, requiring separate optimization for input and subsequent stages. This design avoids that trade-off, allowing three identical stages and a simplified layout process.

\begin{figure}[t]
 		\centering
 		\includegraphics[width=85.0mm]{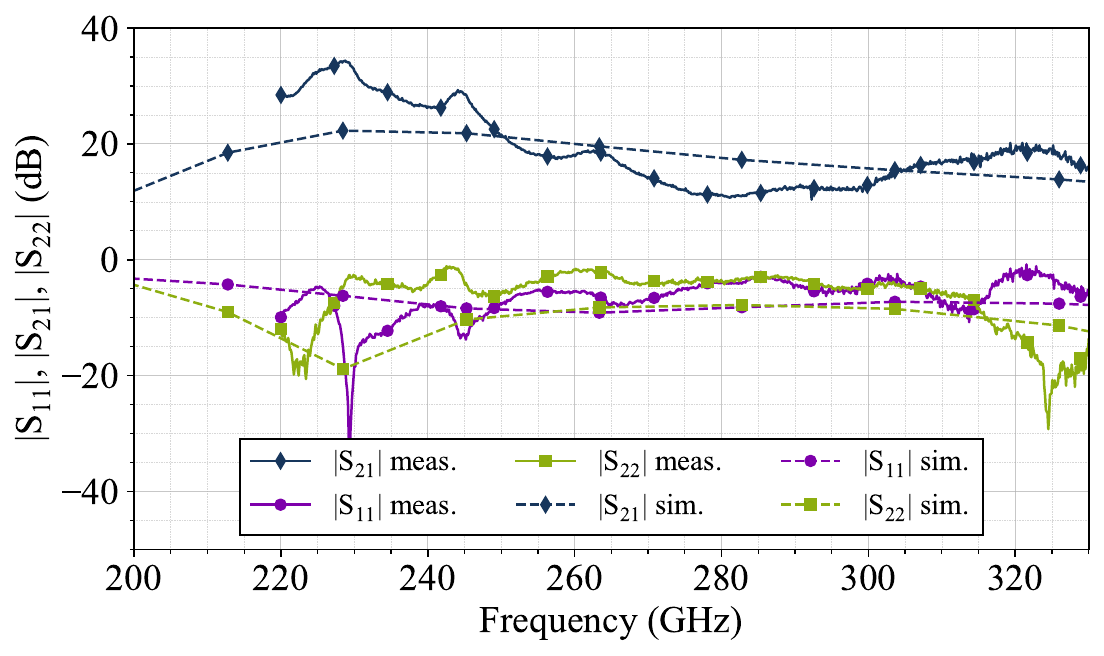}
		\caption{Measured and simulated $S$-parameters of the LNA breakout including pads and balun.}

 		\label{fig:LNA_Sparam}
 		\vspace{-\baselineskip}
 	\end{figure}

The LNA was characterized with a calibrated on-wafer S-parameter measurement of the LNA breakout. For this purpose, a VDI WR3.4 VNAX and a Keysight PNAX were used.
Measurement results are shown in Fig.~\ref{fig:LNA_Sparam}.
Especially at higher frequencies, the LNA exhibits higher gain than simulated, while the gain decreases around the center frequency.
This can be caused by parasitic layout inductances at the base node or capacitances at the emitter node of the cascode stage.
Both elements can lead to gain peaking at the observed frequencies and are difficult to estimate, since no RLC extraction could be performed with the utilized technology design kit.

 	 	\begin{figure}[t]
 		\centering
 		\includegraphics[width=65.0mm]{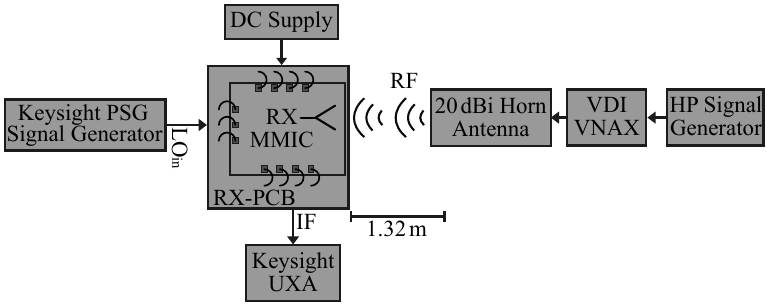}
 		\caption{Setup for the RX free-space characterization. All results are corrected for free-space path loss and TX horn antenna gain. A PTFE lens was not used. The transmit source power was calibrated using a PM5B power meter.}

 		\label{fig:RX_freespace_setup}
 	\end{figure}

		\begin{figure}[t]
 		\centering
 		\includegraphics[width=88.0mm]{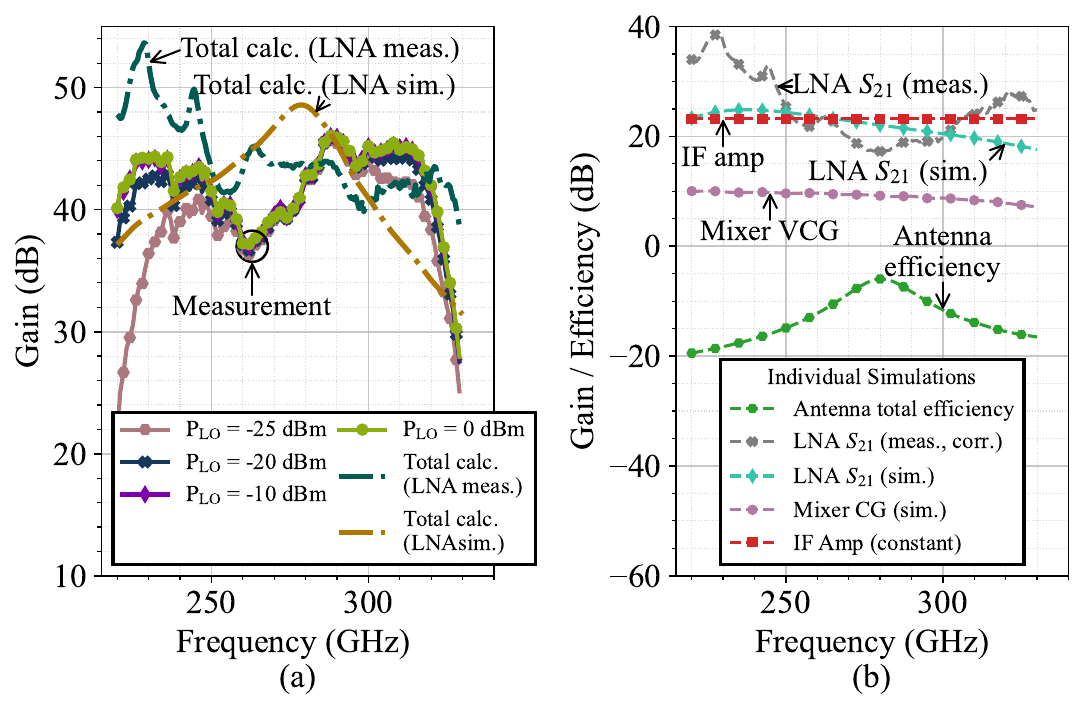}
		\caption{Measured RX conversion gain including antenna losses, LNA, mixer, and IF amplifier, obtained over a 1.34\,m free-space link using a VDI WR3.4 VNAX and a 25\,dBi horn antenna~(a). Free-space path loss, horn antenna gain, and simulated on-chip antenna directivity are de-embedded. The simulated gain contributions of individual components are shown in~(b) and summed to obtain the expected calculated conversion gain.}

 		\label{fig:Meas_RX_Conversion_Gain}
 		\vspace{-\baselineskip}
 	\end{figure}

		\begin{figure}[t]
 		\centering
 		\includegraphics[width=88.0mm]{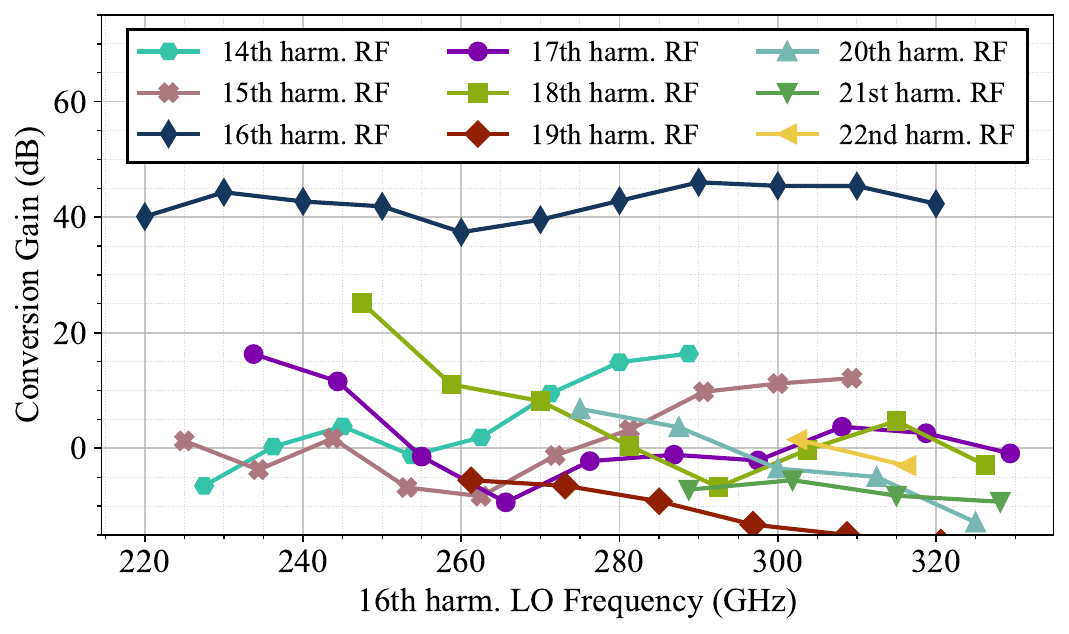}
		\caption{Measured RX conversion gain for different RF harmonic excitations vs.\ 16th harmonic LO frequency, measured over a free-space link.}

 		\label{fig:Meas_RX_Conversion_Gain_Harmonics}
 	\end{figure}

			\begin{figure}[t]
 		\centering
 		\includegraphics[width=88.0mm]{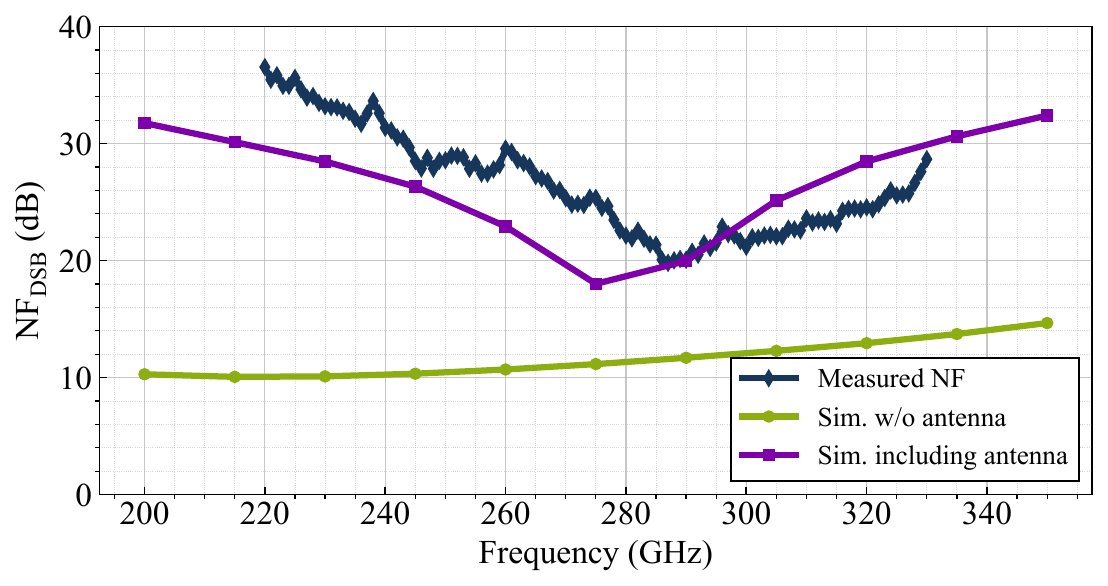}
		\caption{Simulated double-sideband noise figure of the active RX components (LNA and mixer) and gain-method-based NF measurement results of the entire RX system including the antenna. The simulated noise figure including the antenna is calculated from the simulated active component noise figure with the addition of the inverse of the simulated antenna efficiency.}

 		\label{fig:Meas_RX_Noise_Figure}
 	\end{figure}

The overall RX conversion gain was measured over free space with a WR3.4 VNAX as signal source and a Keysight UXA to analyze the IF signal. This measurement setup is shown in Fig.~\ref{fig:RX_freespace_setup}. The corresponding measurement results are shown in Fig.~\ref{fig:Meas_RX_Conversion_Gain}, where the measured conversion gain of the entire RX system is plotted for different LO input powers. To verify the result, a calculated conversion gain based on the individual component measurements is also shown.
The overall conversion gain is the sum of the simulated antenna gain, the measured LNA gain, which was corrected by the back-to-back measured losses of the balun and the pad, the simulated mixer conversion gain and the simulated IF amplifier gain, which is constant at 23.2\,dB.
Analyzing the measurement results, it is notable that the LO input does not have a significant impact on the receiver functionality, as soon as the input power is higher than $-20\,\text{dBm}$, which corresponds to the expected input preamplifier saturation behavior, which was discussed in Section~\ref{Kap:X16}.

The measured conversion gain tracks the LNA resonances seen in the breakout measurement, though with smaller gain peaks. Frequency shifts between measured and calculated values are expected, as antenna-LNA impedance mismatch was neglected in the calculation. A peak conversion gain of $46.2\,\text{dB}$ at $289\,\text{GHz}$ and less than $10\,\text{dB}$ variation from $220\,\text{GHz}$ to $320\,\text{GHz}$ are achieved, driven by the complementary resonance frequencies of the patch antenna and LNA\@.

Although the transmitted harmonics of integrated radar systems are widely characterized, the receiver-side sensitivity to these harmonics remains largely unaddressed. A complete assessment of harmonic-induced artifacts in FMCW radar requires analysis of both sides. Accordingly, the RX conversion gain was measured for multiple harmonic excitations with the LO generated by the proposed $\times 16$ multiplier chain, with results presented in Fig.~\ref{fig:Meas_RX_Conversion_Gain_Harmonics}. 
The measurement setup is identical to that of the previous conversion gain measurement, presented in Fig.~\ref{fig:RX_freespace_setup}, with the RF signal now applied at different harmonics of the LO frequency.
At lower frequencies, the conversion gain of the 17th and 18th harmonics increases, driven by the rising efficiency of the patch antenna. The conversion gain of the 16th harmonic exceeds that of the 18th harmonic by more than 17\,dB at an LO frequency of 247\,GHz. At 289\,GHz, the 16th and 20th harmonic conversion gains differ by more than 29\,dBc.

Noise figure measurements using the gain method were performed to characterize the RX system, as no noise source for the corresponding waveguide band is available. The noise figure was calculated according to~\cite{6178029} from the conversion gain shown in Fig.~\ref{fig:Meas_RX_Conversion_Gain}.
The noise floor was measured with a R\&S FSW at an RBW and VBW of 100\,Hz, averaged over 300 measurements. Fig.~\ref{fig:Meas_RX_Noise_Figure} shows the measured noise figure alongside the simulated noise figure of the active receiver components excluding the antenna and a simulated noise figure including the antenna. To account for the antenna noise contribution, the simulated inverse of the antenna efficiency was added to the noise figure of the active components. As both quantities refer to a 100\,$\Omega$ reference impedance, frequency-dependent mismatch was in parts neglected. The minimum measured noise figure of approximately 20\,dB at 285\,GHz agrees well with simulation. 

\section{Radar System}
\label{Kapitel:System}
The proposed RX and TX MMICs were placed on two separate RO4350B PCBs which include LDOs to provide the 3.3\,V supply voltage for the MMICs, as well as connectors for the external LO signals.
The RX module also includes an IF amplifier to provide high impedance at the mixer output and to amplify the IF signal. On both modules a collimating PTFE lens can be mounted to increase the antenna gain.
The MMIC is then placed in the focal point of the lens, within a small cutout at the bottom of the lens to give space for the  bond wires~\cite{10959113}. 

\begin{figure}[t]
 		\centering
 		\includegraphics[width=88.0mm]{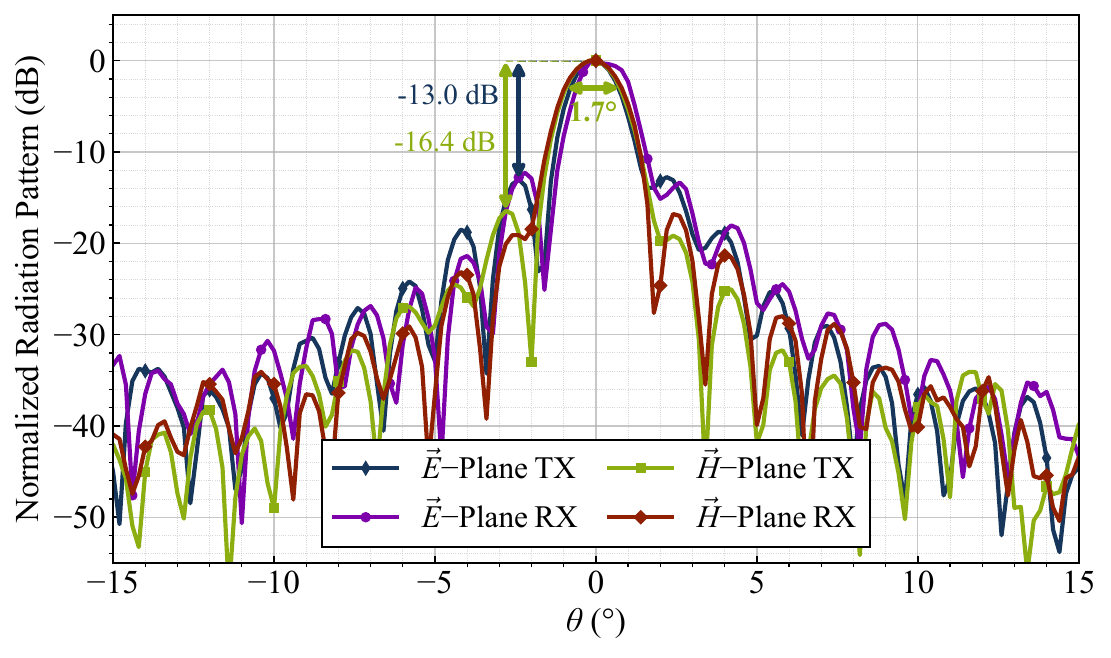}
		\caption{Measured radiation pattern for RX and TX modules with lens in $\vec{E}$ and $\vec{H}$-Plane. The 3\,dB beamwidth is around 1.7\,° and the sidelobe levels are between -13 and -16.4\,dB.	}

 		\label{fig:Directivity}
 	\end{figure}

While the on-chip patch antenna alone provides a wide beamwidth, the combination with the lens significantly increases the antenna gain and narrows the beamwidth. 
The measured radiation pattern of the TX and RX module with lens is shown in Fig.~\ref{fig:Directivity}. The 3\,dB beamwidth is around 1.7\,° and the sidelobe levels are less than -13\,dB.
The TX and RX modules were positioned opposite each other with a separation exceeding 2\,m.
The TX module was operated in mono-frequency mode, providing a continuous-wave signal that was down-converted by the RX module to a fixed IF\@.
To characterize each module individually, either the TX or the RX was rotated horizontally and vertically by precise stepper motors, while the complementary module remained stationary and was operated without a lens.
The resulting IF amplitude variations were recorded.
Amplitude and angle of the measured radiation pattern are normalized to the maximum measured value.

\begin{figure}[t]
 		\centering
 		\includegraphics[width=88.0mm]{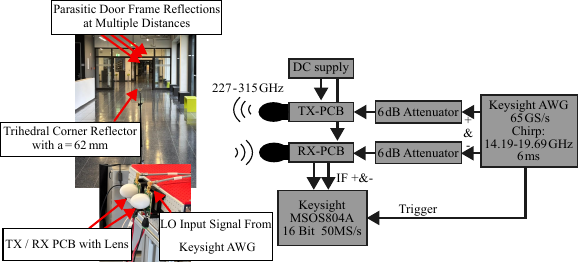}
		\caption{Measurement setup for FMCW range measurements. RX and TX modules are placed quasi monostatic on a common fixture. The FMCW chirp is generated by a Keysight AWG. The IF signal is sampled by a Keysight MSOS804A oscilloscope and processed by a PC.	}

 		\label{fig:FMCW_Range_Setup}
 		\vspace{-\baselineskip}
 	\end{figure}
\begin{figure}[t]
 		\centering
 		\includegraphics[width=88.0mm]{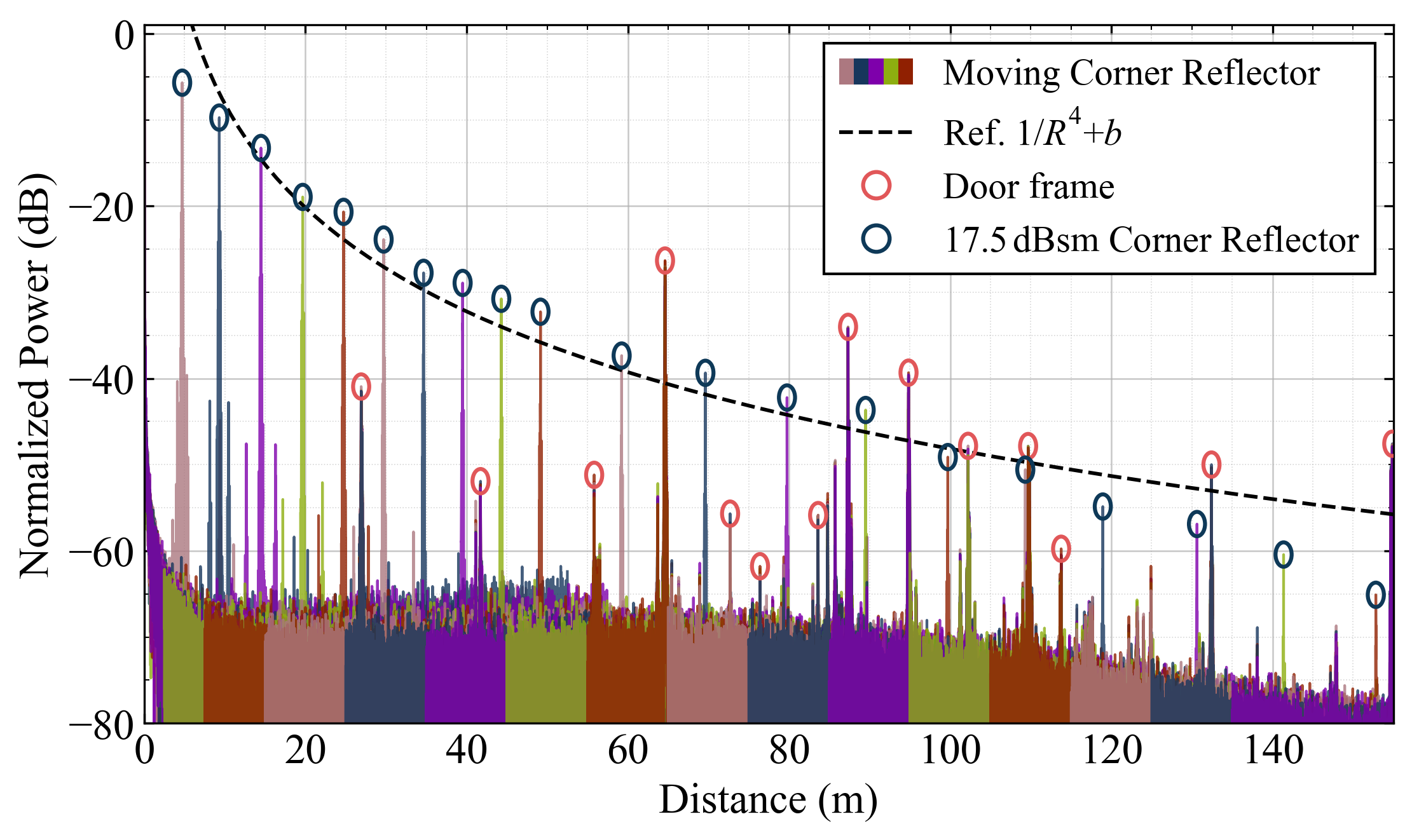}
		\caption{FMCW range measurements with a moving corner reflector. Measurements were taken inside a hallway. Peaks that clearly correspond to door frame reflections are marked with red circles. Peaks that are caused by the corner reflector are marked with blue circles. 	}

 		\label{fig:FMCW_Range}
 		\vspace{-\baselineskip}
 	\end{figure}

\renewcommand{\arraystretch}{1.25}

\begin{table*}[]
	\centering
	\caption{Comparison of state-of-the-art J-band radar transmitters and receivers}
	\label{Tabelle_Comparison}
\begin{tabular}{ccccccccc}
\toprule
\textbf{Study}                                                                     & \cite{10147390}                & \cite{10663766}     & \cite{10934734}     & \cite{10025412}                & \cite{Ziegler_Bellenberg_2025} & \cite{9906590}        & \cite{GEMICPRTX}              & \textbf{This work}                                                     \\
\midrule
\textbf{Technology}                                                                & 130\,nm SiGe                   & 130\,nm SiGe        & 65\,nm CMOS         & 130\,nm SiGe                   & 90\,nm SiGe                    & 130\,nm SiGe          & 90\,nm SiGe                    & \textbf{130\,nm SiGe}                                                  \\[3pt]
\hline
\textbf{f$_{\textup{t}}$/f$_\textup{max}$\,(GHz)}                                  & 300/500                        & 250/370             & -                   & 300/500                        & 300/520                        & 300/500               & 300/530                        & \textbf{500/610}                                                       \\[3pt]
\hline
\textbf{Type}                                                                      & $\times$18 TRX                 & \begin{tabular}[c]{@{}c@{}}VCO+\\ $\times$2 TRX\end{tabular}   & $\times$16 TX/RX    & $\times$12 TX/RX               & \begin{tabular}[c]{@{}c@{}}VCO+\\ $\times$4 TRX\end{tabular}              & $\times$8 TRX         & $\times$16 TX                  & \textbf{$\times$16 TX/RX}                                              \\[3pt]
\hline
\textbf{Antenna}                                                                   & \begin{tabular}[c]{@{}c@{}}On-Chip +\\ Silicon Lens\end{tabular} & On-Chip & On-Chip & \begin{tabular}[c]{@{}c@{}}On-Chip\,+\\ Silicon Lens\end{tabular} & \begin{tabular}[c]{@{}c@{}}On-Chip\,+\\ PTFE Lens\end{tabular} & On-Chip & \begin{tabular}[c]{@{}c@{}}On-Chip+\\ PTFE Lens\end{tabular} & \textbf{\begin{tabular}[c]{@{}c@{}}On-Chip +\\ PTFE Lens\end{tabular}} \\[3pt]\hline
\textbf{f$_\textup{c}$\,(GHz)}                                                     & 242                            & 235                 & 254                 & 226$^a$                        & 285                            & 256                   & 270                            & \textbf{292}                                                           \\[3pt]\hline
\textbf{\begin{tabular}[c]{@{}c@{}}BW$_\textup{-3dB}$/TR\\ (GHz)\end{tabular}}    & 50/-                           & -/33                & 30/-                & 7.5$^a$/-                      & -/90                           & 65/-                  & 23/-                           & \textbf{32/-}                                                          \\[3pt]
\hline
\textbf{\begin{tabular}[c]{@{}c@{}}EIRP\,(dBm)\\ with / w/o lens\end{tabular}}     & 27/-                           & -/1.5               & -/-2                & 18/10                          & 9.2/-                          & -5.4$^b$              & 31/1.7                         & \textbf{41/8.8}                                                        \\[3pt]
\hline
\textbf{\begin{tabular}[c]{@{}c@{}}TX HRR\\ at f$_\textup{c}$ (dBc)\end{tabular}} & 34$^c$                         & -                   & -                   & -                              & -                              & -                     & 30$^d$/35$^c$                  & \textbf{50$^c$}                                                        \\[3pt]\hline
\textbf{RX CG\,(dB)}                                                               & 16.2                           & 12$^e$              & 4$^a$               & 48$^h$                         & 1.22$^e$                       & 10.4                  & -                              & \textbf{43.3$^h$\,/\,23.1}                                                             \\[3pt]
\hline
\textbf{NF$_\textup{DSB}$\,(dB)}                                                   & 15.7$^f$                       & 14$^{e,f}$          & 25$^g$              & 6$^f$                          & 22.5$^{a,e,i}$                 & 20.5$^{e,f}$          & -                              & \textbf{20$^g$}                                                             \\[3pt]\hline
\textbf{Area\,(mm$^2$)}                                                            & 2.7                            & 2.8                 & 1.4\,/\,1.8$^j$         & 1.82\,/\,1.54$^j$                  & 3.51                           & 3.3                   & 3.51                           & \textbf{1.66\,/\,1.8$^j$}                                                  \\[3pt]\hline
\textbf{P$_\textup{DC}$\,(W)}                                                      & 0.77                           & 0.85                & 0.24\,/\,0.29$^j$       & 0.64\,/\,0.5$^j$                  & 0.7                            & 0.3                   & 1.5                            & \textbf{1.87\,/\,0.78$^j$}                                                        \\
\bottomrule
\end{tabular}

\footnotesize{$^a$\,estimated from graph,\;
$^b$\,P$_{\textup{TX}}$,\;
$^c$\,to J-band harmonics,\;
$^d$\,to D- and J-band harmonics,\;
$^e$\,simulated,\;
$^f$\,exclusive of antenna,\;
$^g$\,inclusive of antenna,\;
$^h$\,inclusive of IF amplifier,\;
$^i$\,SSB or DSB unclear
$^j$\,inclusive TX\,/\,RX MMIC only,\;}
\end{table*}

To verify the functionality of the entire radar chipset, FMCW range measurements were performed with the setup shown in Fig.~\ref{fig:FMCW_Range_Setup}. The TX and RX modules were placed in a quasi-monostatic configuration, and a Keysight AWG with 65\,GS/s was used to generate the FMCW chirp as the LO input signal for the frontend modules over a frequency range from 14.19 to 19.69\,GHz, corresponding to a transmitted chirp spanning from 227 to 315\,GHz. The IF signal was sampled by a Keysight MSOS804A oscilloscope with 16-bit vertical resolution. A single chirp with a ramp time of 6\,ms was processed on a PC. No additional signal processing steps, such as zero-padding or window functions, were applied. To mitigate the influence of impedance mismatches and reflections of the AWG chirp signal within the LO path, 6\,dB attenuators were placed directly at the LO input of the TX and RX modules. The measurement was conducted in a hallway using a moving corner reflector as the target, which provides a radar cross section of approximately 17.5\,dBsm at 280\,GHz.
Fig.~\ref{fig:FMCW_Range} shows the range measurement results for multiple target distances. Due to the high system EIRP, the corner reflector can be clearly identified in FMCW operation up to a range of 150\,m. The IF power decreases significantly for distances above 100\,m, where the corresponding IF frequency of 9.8\,MHz exceeds the 3\,dB corner frequency of the IF amplifier at 7.5\,MHz (77\,m). For shorter distances, the IF power follows the expected $1/r^4$ decrease.
Within the hallway, multiple reflections from the door frames are visible, which are indicated by circles in the plot.
Ghost targets visible close to the actual target, particularly at shorter distances, are most likely attributable to the AWG chirp signal, as they are absent when other chirp generators are employed.

\section{Conclusion}
\label{Kapitel:Conclusion}
This work presented a J-band radar chipset achieving an EIRP of 41\,dBm at 292\,GHz, exceeding previously reported values for J-band radar transmitters of 27\,dBm in~\cite{10147390} and 31\,dBm in~\cite{GEMICPRTX}, both obtained with a lens. A detailed comparison of the proposed TX and RX MMICs with state-of-the-art J-band radar frontends is provided in Table~\ref{Tabelle_Comparison}.
Operation without a lens is also feasible, providing a wide field of view at an EIRP of 8.8\,dBm.
Although the patch antenna limits the 3\,dB bandwidth to 32\,GHz, a broad 10\,dB bandwidth of 88\,GHz is achieved.
The integrated $\times 16$ frequency multiplier permits a low-frequency LO source, enabling a scalable radar setup with multiple TX and RX modules.
High multiplication orders typically introduce strong harmonic spurs near the carrier frequency.
Nevertheless, a maximum radiated harmonic rejection of 50\,dBc, remaining above 21.6\,dBc from 227 to 316\,GHz, is achieved through filter stages in the multiplier chain together with the filtering response of the patch antenna.
For the RX module, a conversion gain of 43\,dB was achieved, exploiting the advantages of multistatic systems. The impact of undesired received harmonics was also evaluated, revealing a conversion-gain difference of approximately 25\,dB between the 16th and 14th harmonic at the center frequency.
The noise figure of the RX system was validated using the gain method, yielding a minimum of approximately 20\,dB at 285\,GHz, which corresponds to the state of the art.
A direct comparison is challenging due to differences in measurement setups, with some works including the antenna~\cite{10934734} and others excluding it~\cite{10147390,10663766}.
For the final system verification, FMCW radar measurements were performed over distances of up to 150\,m, demonstrating the long-range capability of the proposed chipset despite the high center frequency of 292\,GHz.

		\bibliographystyle{IEEEtran}
\bibliography{IEEEexample}
\begin{IEEEbiography}[{\includegraphics[width=1in,height=1.25in,clip,keepaspectratio]{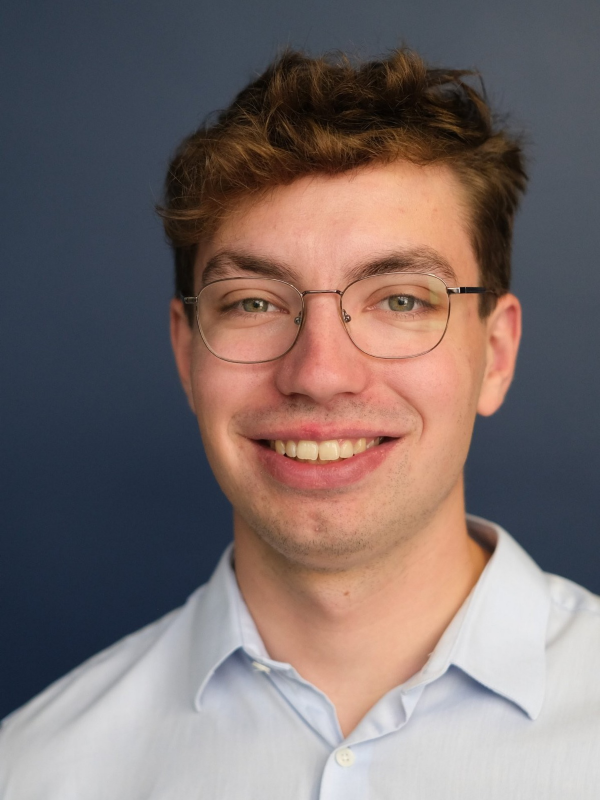}}]{Stephan Hauptmeier}  (Graduate Student Member, IEEE) received the B.Sc. and M.Sc. degrees in electrical engineering from Ruhr University Bochum, Bochum, Germany, in 2022 and 2025, where he is currently pursuing his Ph.D. degree. 
Since 2021, he has been a Research Assistant with the Institute of Integrated Systems, Ruhr University Bochum. His current research interests include integrated circuits for radar sensors in the sub-mm-wave range.
Mr. Hauptmeier received the best paper award at the German Microwave Conference in 2026.

\end{IEEEbiography}

\begin{IEEEbiography}[{\includegraphics[width=1in,height=1.25in,clip,keepaspectratio]{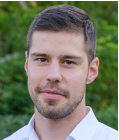}}]{Kennet Braasch} (Member, IEEE) became a Member in 2022. This author received the B.Sc. and M.Sc. in electrical engineering and information technology from the Kiel University, Germany in 2018 and 2020, respectively.
Since 2020, he has been working as a Research Associate and pursuing the Dr.-Ing. degree as a member of the Chair of Microwave Engineering with the Institute of Electrical Engineering and Information Technology at Kiel University. His current research topics include radar technology, antennas, passive components, and particle measurements with various radar sensors.

\end{IEEEbiography}

\begin{IEEEbiography}[{\includegraphics[width=1in,height=1.25in,clip,keepaspectratio]{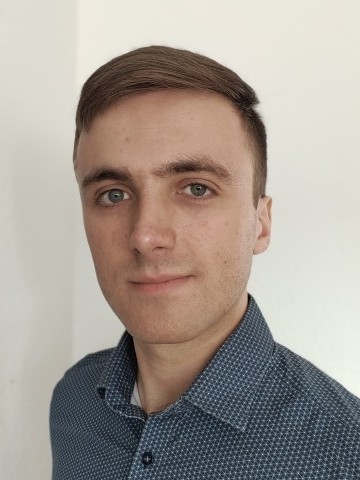}}]{Till Ziegler-Bellenberg} (Graduate Student Member, IEEE)  
received the B.Sc. and
M.Sc. degrees in electrical engineering and
information technology from Ruhr-University
Bochum, Bochum, Germany, in 2019 and 2022,
respectively, where he is currently pursuing the
Ph.D. degree. Since 2023, he has been with the
department of Integrated Circuits, Fraunhofer
Institute for High Frequency Physics and Radar
Techniques FHR, Wachtberg, Germany. His
research interests include the design of integrated SiGe circuits and system concepts for
communication and sensing above 200 GHz.
\end{IEEEbiography}

\begin{IEEEbiography}[{\includegraphics[width=1in,height=1.25in,clip,keepaspectratio]{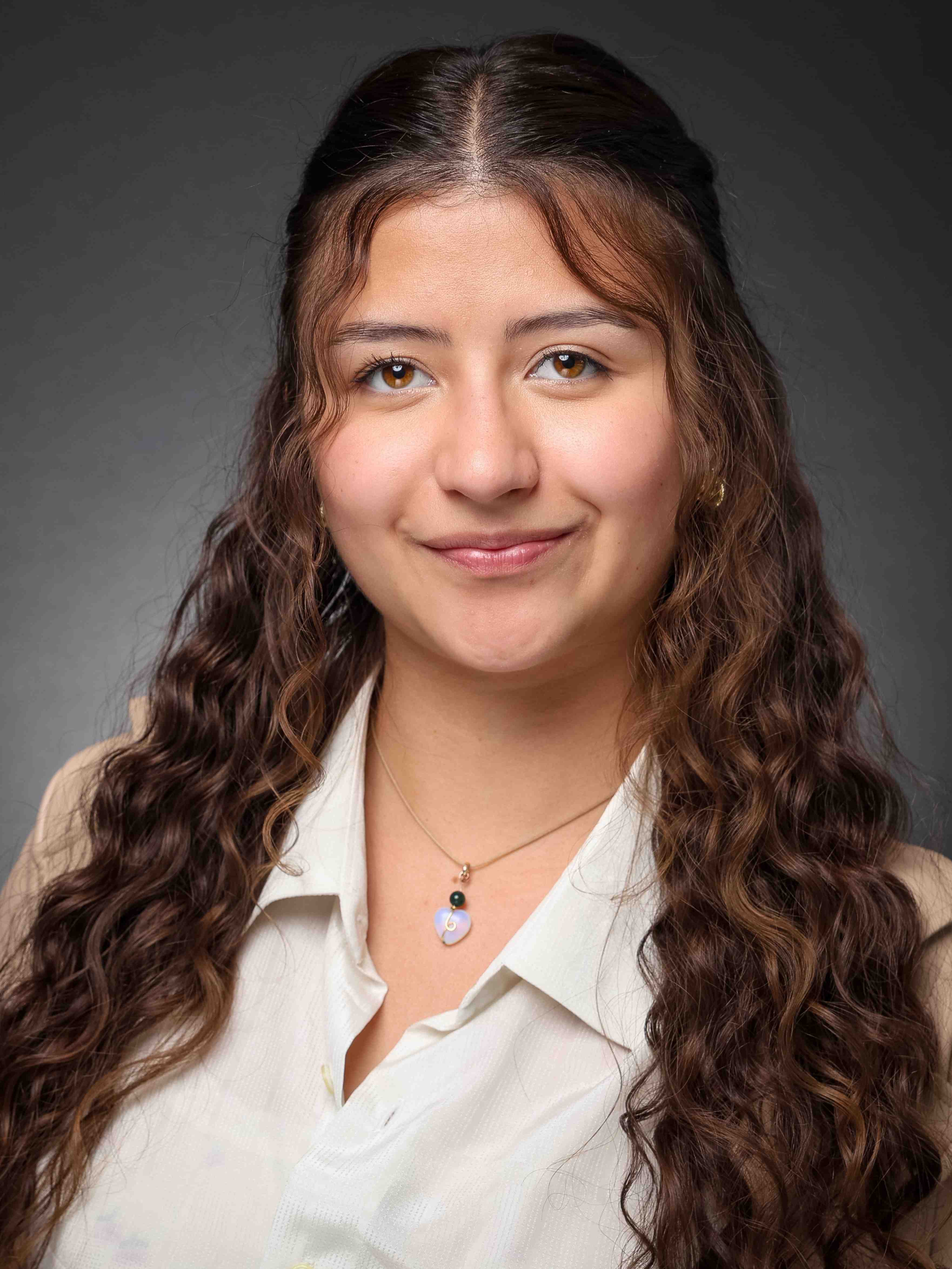}}]{Diana P. Cortés N.}  
was born in Bogotá, Colombia, in 2002. She is currently pursuing her M.Sc. degree in Electrical Engineering and Information Technology at the Ruhr University Bochum, Germany. She participated in a double degree program between the Ruhr University Bochum, and the Universidad Nacional de Colombia, earning a bachelor degree from both universities. From 2022 to 2024, she was a Research Assistant at the research group of High Frequency Electronics and Telecommunications, focusing on a project that aims to safeguard biodiversity by remotely monitoring the Amazon region. Since 2024 she has been a Research Assistant with the Institute of Integrated Systems, Ruhr University Bochum, with a focus on printed circuit board design for radar systems.
\end{IEEEbiography}

\begin{IEEEbiography}[{\includegraphics[width=1in,height=1.25in,clip,keepaspectratio]{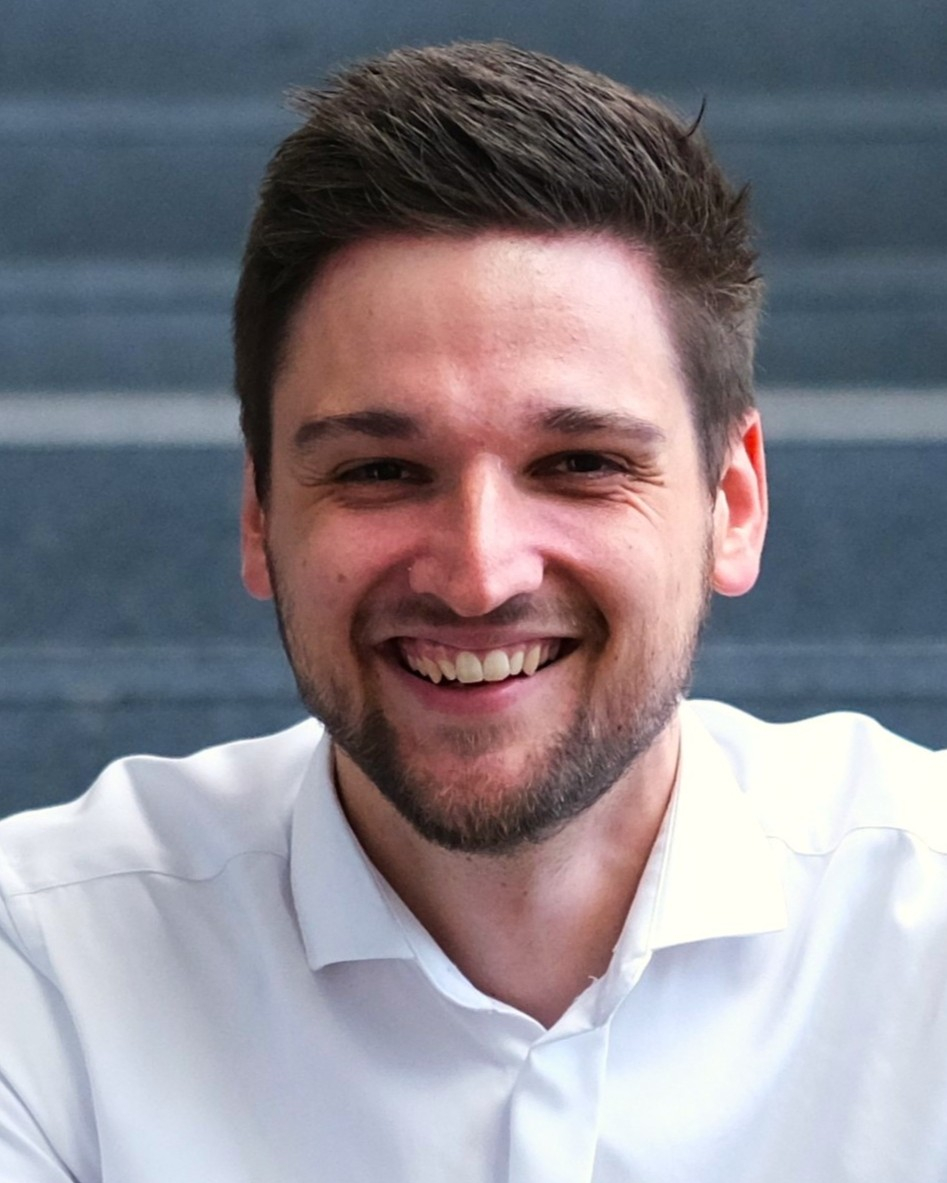}}]{Tobias T. Braun}  
	(Member, IEEE) was born in Duesseldorf, Germany, in 1996. He received the B.Sc. and M.Sc. degrees from TU Dortmund, Dortmund, Germany, in 2016 and 2019, respectively, and the Ph.D. degree from Ruhr University Bochum, Bochum, Germany in 2024, all in electrical engineering and information technology.

	From 2019 to 2025, he was a Research Assistant at the Institute of Integrated Systems at Ruhr University Bochum, where he has since become the Institute’s Chief Engineer. His current research interests include circuit and system design for mmWaves and (sub-)THz frequencies with a focus on automotive applications, low noise synthesizers, space observation, and nonlinear radar systems. 

	Tobias T. Braun was the recipient of the EuMIC Young Engineer Prize from European Microwave Week in 2021, Honorable Mentions of the 3MT competition at the International Microwave Symposium 2022, the 2024 International Journal of Microwave and Wireless Technologies Best Paper Award, 2nd Place of the Student Paper Competition at Radio \& Wireless Week 2025, and of the 2025 VDE ITG Dissertationspreis for his Ph.D. thesis. 
\end{IEEEbiography}

\begin{IEEEbiography}[{\includegraphics[width=1in,height=1.25in,clip,keepaspectratio]{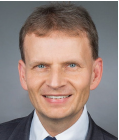}}]{Michael Höft}
(Senior Member, IEEE) was born in Lübeck, Germany, in 1972. The author received the Dipl.-Ing. degree in electrical engineering and the Dr.-Ing. degree from the Hamburg University of Technology, Hamburg, Germany, in 1997 and 2002, respectively.

 From 2002 to 2013, he joined the Communications Laboratory, European Technology Center, Panasonic Industrial Devices Europe GmbH, Lüneburg, Germany. There he was first a Research Engineer and then a Team Leader, where he had been engaged in the research and development of microwave circuitry and components, particularly filters for cellular radio communications. Then he was at the same organization from 2010 to 2013 a Group Leader for the research and development of sensor and network devices. Since October 2013 he has been a Full Professor at Kiel University, Kiel, Germany, in the Faculty of Engineering, where he heads the Chair for Microwave Engineering of the Institute of Electrical and Information Engineering. His research interests include active and passive microwave components, (sub-)millimeter-wave quasi-optical techniques and circuitry, microwave and field measurement techniques, microwave filters, microwave sensors, and magnetic field sensors as well as related applications. Dr. Höft is a member of the European Microwave Association (EuMA), the Association of German Engineers (VDI), a member of the German Institute of Electrical Engineers (VDE), and a Senior Member of the Institute of Electrical and Electronics Engineers (IEEE).
\end{IEEEbiography}

\begin{IEEEbiography}[{\includegraphics[width=1in,height=1.25in,clip,keepaspectratio]{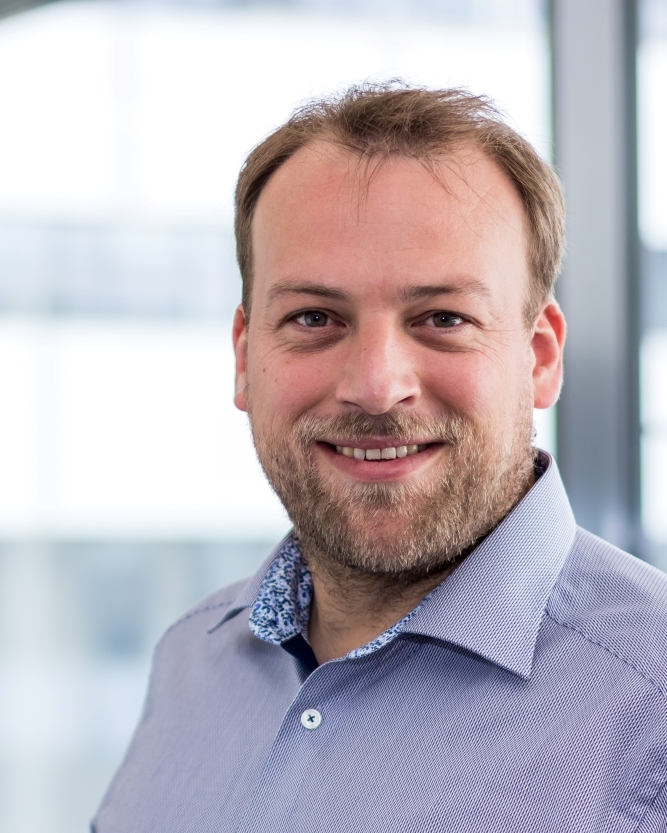}}]{Nils Pohl}
	(Fellow, IEEE) received the Dipl.-Ing. and Dr.-Ing. degrees in electrical engineering from Ruhr University Bochum, Bochum, Germany, in 2005 and 2010, respectively.
From 2006 to 2011, he was a Research Assistant and subsequently from 2011 an Assistant Professor with Ruhr University Bochum. In 2013, he became the Head of the Department of mm-wave radar and high frequency sensors with the Fraunhofer FHR, Wachtberg, Germany. Since 2016, he has been a full professor of integrated systems at Ruhr University Bochum, while also maintaining a part-time affiliation with Fraunhofer FHR. He is involved in several third-party-funded research projects and is a co-founder of the start-up 2Pi-Labs. He is currently supervisor of 27 PhD students and several PostDocs. He has authored or coauthored more than 300 scientific papers and has issued several patents. His current research interests include ultra-wideband mm-wave radar, design, and optimization of mm-wave integrated SiGe circuits and system concepts with frequencies up to 500 GHz, frequency synthesis, and antennas.
He is a member of IEEE, VDE, ITG, EUMA, and URSI. He is actively involved in the scientific community, having served on the program committees of IMS, EUMW, RWW, SIRF, BCICTS, GEMIC, IMWS-AMP, and EUCAP. He is a reviewer for several IEEE journals, has been a Guest Editor for TMTT, and is currently Associate Editor-in-Chief for Transactions on Radar Systems. He served as chair of the MTT TC 24 “Microwave/mm-Wave Radar, Sensing, and Array Systems Committee” and is currently the spokesman of the microwave section in the German VDE/ITG.
He received the Karl-Arnold Award from the North Rhine-Westphalian Academy of Sciences, Humanities and the Arts in 2013, the International IHP “Wolfgang Mehr” Fellowship Award for research in high-frequency electronics in 2017, the IEEE MTT Outstanding Young Engineer Award in 2018, and the VDE-ITG Award in 2023. He was also a co-recipient of the Best Demo Award at RWW 2015, Best Student Paper Awards at Radar 2020, RWW 2021, and EuMIC 2021, and Best Paper Awards at EuMIC 2012, AWPL 2022, SENSL 2023, IJMWT 2025, and GEMIC 2026.
\end{IEEEbiography}

\vfill

\end{document}